\documentclass[fleqn,12pt]{wlscirep}
\usepackage[utf8]{inputenc}
\usepackage[T1]{fontenc}
\usepackage[numbers,super]{natbib}
\usepackage{caption}
\usepackage{subcaption}
\usepackage{float}

\newcommand{\farcs}{\mbox{\ensuremath{.\!\!^{\prime\prime}}}}
\newcommand{\apjl}{Astrophys. J. Lett.}   
\newcommand{\apjs}{Astrophys. J. Suppl. Ser.}   
\newcommand{\aap}{Astron. Astrophys.}   

\newcommand{\xmm}{\hbox{\hbox{XMM-Newton }\/}}

\newcommand{\ion}[2]{\text{#1\,\textsc{#2}}}
\newcommand{\sersic}{S\'{e}rsic}

\title{\large Discovery of two little red dots transitioning into quasars}

\author[1,2,6]{Shuqi Fu}
\author[1,2,6]{Zijian Zhang}
\author[1,2]{Danyang Jiang}
\author[1,2]{Jie Chen}
\author[1,2,*]{Linhua Jiang}
\author[1,2]{Luis C. Ho}
\author[1,2]{Kohei Inayoshi}
\author[1,2]{Kaiyuan Chen}
\author[3]{Jianwei Lyu}
\author[4]{Fengwu Sun}
\author[5]{Feige Wang}
\author[5]{Jinyi Yang}

\affil[1]{Department of Astronomy, School of Physics, Peking University, Beijing 100871, China}
\affil[2]{Kavli Institute for Astronomy and Astrophysics, Peking University, Beijing 100871, China}
\affil[3]{Steward Observatory, University of Arizona, 933 N. Cherry Ave., Tucson, AZ 85721, USA}
\affil[4]{Center for Astrophysics $|$ Harvard \& Smithsonian, 60 Garden St., Cambridge, MA 02138, USA}
\affil[5]{Department of Astronomy, University of Michigan, 1085 S. University Ave., Ann Arbor, MI 48109, USA}
\affil[6]{These authors contributed equally: Shuqi Fu, Zijian Zhang}
\affil[*]{e-mail: linhua.jiang@pku.edu.cn}

\makeatletter
\renewcommand{\@maketitle}{%
{%
\thispagestyle{empty}%
\vskip-36pt%
{\raggedright\sffamily\bfseries\fontsize{20}{25}\selectfont \@title\par}%
\vskip10pt
{\raggedright\sffamily\fontsize{12}{16}\selectfont  \@author\par}
\vskip25pt%
}%
}%
\makeatother

\begin{document}

\flushbottom
\maketitle
\noindent \textbf{James Webb Space Telescope (JWST) has uncovered a new population of compact objects that show a unique V-shaped spectral energy distribution (SED) in the UV and optical wavelength range. These so-called ``little red dots'' (LRDs) often exhibit broad Balmer emission lines, indicative of the presence of active galactic nuclei (AGNs). They generally lack detection of X-ray, radio, and mid-IR radiation, which is fundamentally different from typical AGNs. Various models, including super-Eddington-accreting black holes enshrouded in dense and dust-poor gas, have been proposed to explain these features. However, the nature of LRDs remains debated, and their evolutionary fate is unclear. Here we report two unusual LRDs at redshift $z = 2.868$ and $2.925$ that are in a transitional phase towards typical AGNs. Their V-shaped SEDs, compact optical morphology, and broad emission lines satisfy the defining criteria of LRDs. On the other hand, they exhibit intense X-ray, radio, and mid-IR radiation that is much stronger than previously known LRDs. These hybrid properties suggest that the dense gas envelope around their central black holes is dispersing, allowing high-energy photons and radio emission to escape. Meanwhile, a dust torus is forming. This finding provides a direct insight into the nature of LRDs and indicates that at least some LRDs will evolve into AGNs or quasars at later times.} 

\medskip

LRDs are characterized by their distinctive V-shaped SED that consists of a red optical continuum and a blue UV continuum\citep{Furtak2023,Matthee2024} with a turnover point consistently located near the Balmer limit\citep{2024arXiv241103424S}. 
They have point-like rest-frame optical morphology with a typical effective radius $R_{\rm e} \sim 100$ pc\citep{Kocevski2025,2024RNAAS...8..207G} or less \citep{Furtak2023,Furtak2025b}, and more than $70\%$ of them show broad Balmer emission lines \cite{Greene2024,Matthee2024,Zhang2025b}. These observational features provide strong evidence for the presence of accreting supermassive black holes (SMBHs). However, some key observational signatures of typical AGNs are weak or absent from LRDs, including X-ray emission from hot corona\cite{Maiolino_JWSTChandra,Yue2024}, mid-IR emission from hot dust\cite{Akins2024a_cos,Pablo2024_smilemiri}, and radio emission\citep{2024arXiv241204224M}. In addition, Balmer absorption is observed in a large fraction ($\geq 20\%$) of LRD spectra\citep{2024arXiv240717570L,2025arXiv250403551J}, supporting a scenario in which a super-Eddington SMBH is enshrouded by very dense gas \citep{Kido2025,2025ApJ...980L..27I,2025arXiv250113082J}. Although most LRDs do not show significant UV–optical variability \citep{2024arXiv240704777K,2025ApJ...985..119Z}, a small fraction of them display variability that is consistent with actively accreting BHs\citep{Furtak2025b,Inayoshi2024_xrayweak,Ji2025a,2025ApJ...985..119Z}.
While the physical mechanism underlying their characteristic properties remains poorly understood, little is known about their evolutionary pathway as well. The LRD occurrence rate follows a log-normal distribution that peaks at $z\sim6$ and declines sharply towards lower redshifts \citep{Kocevski2025,Ma2025,Tanaka2025}, so LRDs may represent an early and short-lived phase of SMBH growth \citep{Inayoshi2025_firstactivity}. The rapid decline toward later times indicates that LRDs do not persist as a distinct population. It is unclear whether LRDs evolve into typical AGNs, and if so, how this transition occurs. 

\medskip

\begin{figure}[H]
    
    \begin{subfigure}{\linewidth}
    \centering
    \includegraphics[width=\linewidth]{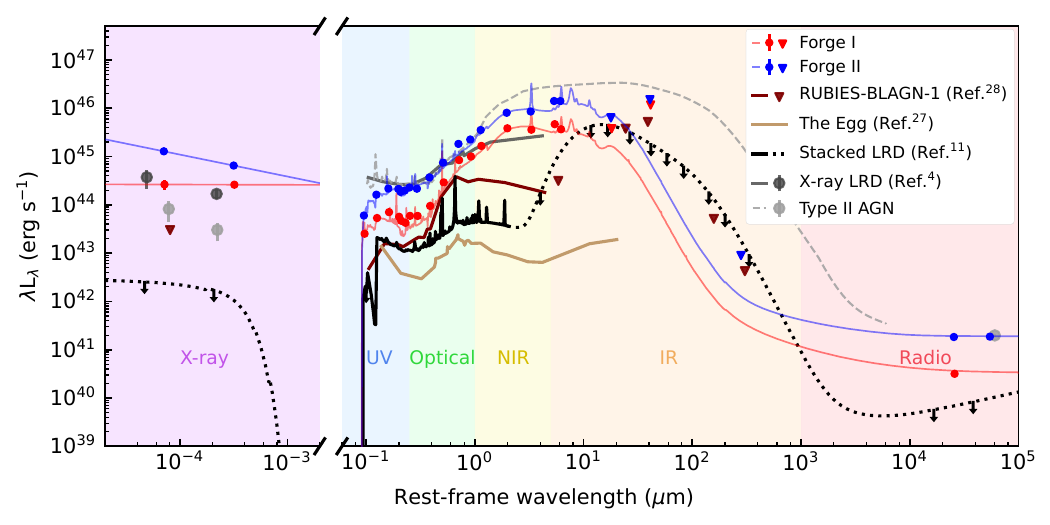}
    \end{subfigure}

    \begin{subfigure}[b]{0.49\linewidth}
        \centering
        \includegraphics[width=\linewidth]{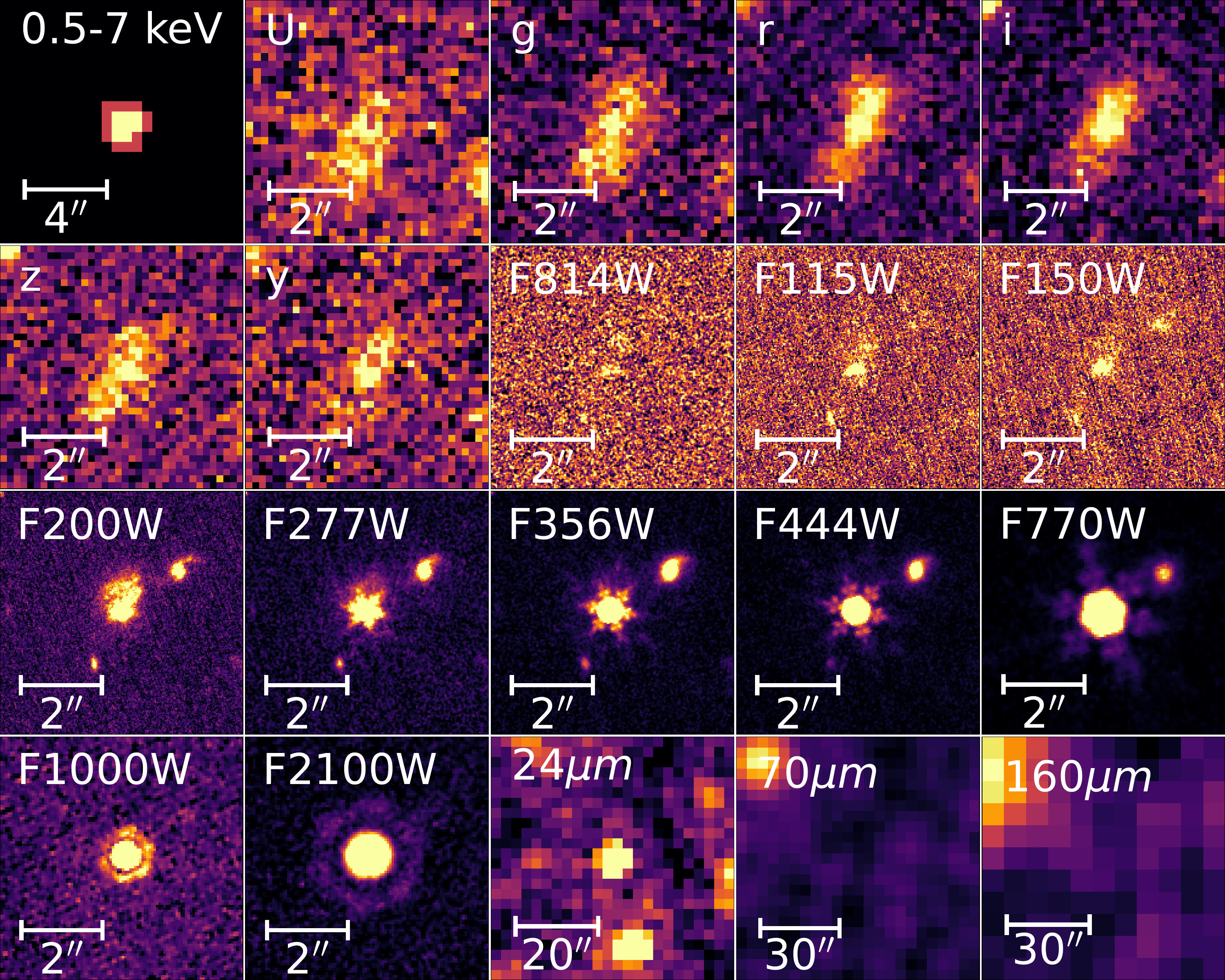}
    \end{subfigure}
    \hfill
    \begin{subfigure}[b]{0.49\linewidth}
        \centering
        \includegraphics[width=\linewidth]{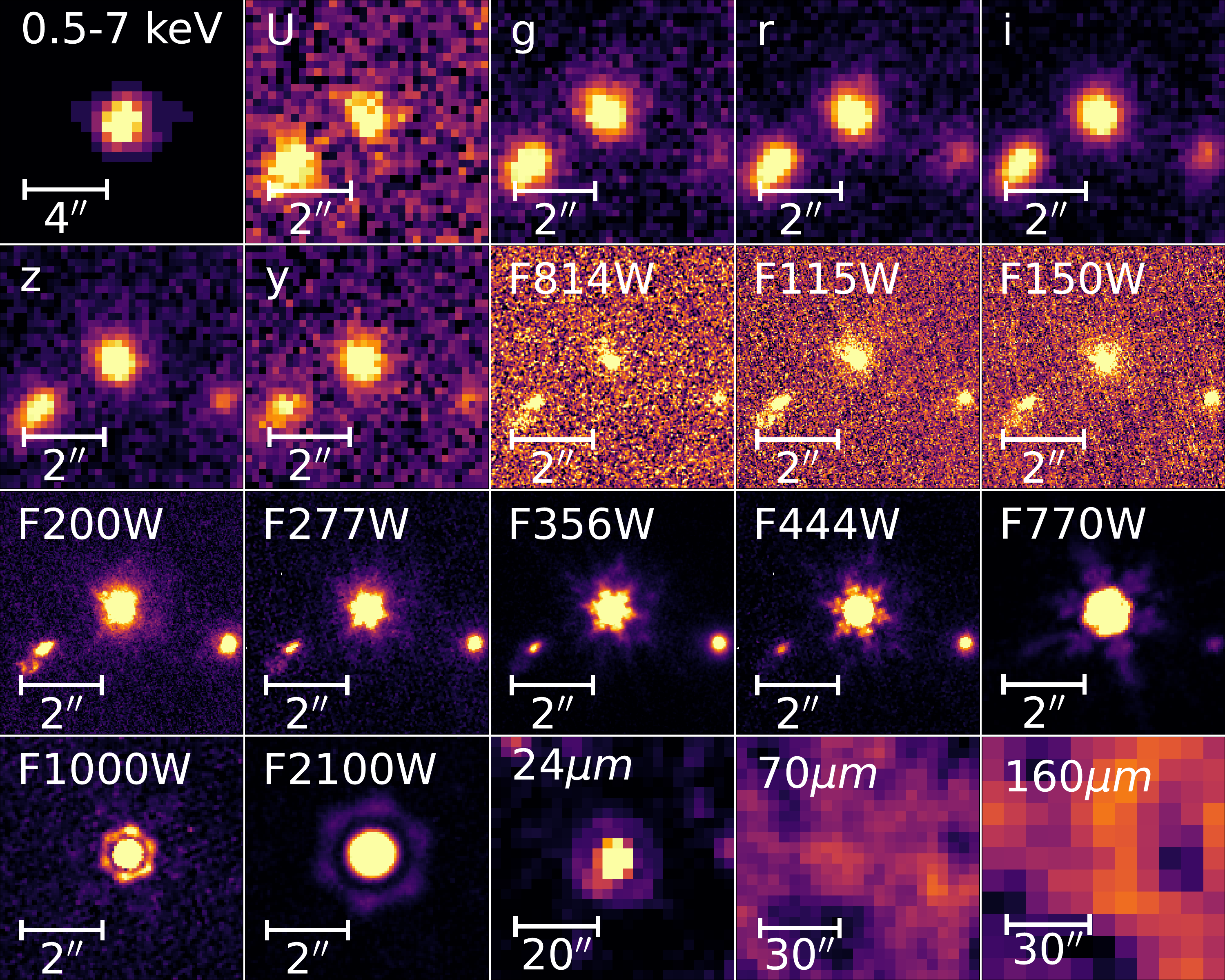}
    \end{subfigure}
    \caption{SEDs and multi-band images of the \textit{Forges}. \textbf{Upper:} The X-ray to radio SEDs of the \textit{Forges}. 
    For reference, we also show the SEDs of other sources: 
    a typical $z = 3.1$ LRD (RUBIES-BLAGN-1\cite{Wang2025}), a local LRD counterpart (\textit{the Egg}\cite{Lin2025_locallrd}), the stacked maximal COSMOS-Web LRD SED where the dotted line represents upper limits\cite{Akins2024a_cos}, another X-ray detected $z=4.66$ LRD in COSMOS field\cite{Kocevski2025}, and a highly obscured Type II AGN\cite{2006ApJ...642..673P}. \textbf{Lower:} Cutout images of the \textit{Forge I} (left) and \textit{Forge II} (right) spanning from X-ray to far-IR, including Chandra X-ray (0.5–7 keV), CFHT $U$, Subaru $grizy$, HST ACS F814W, JWST NIRCam F115W, F150W, F200W, F277W, F356W, F444W, JWST MIRI F770W, F1000W, F2100W, and Spitzer MIPS 24, 70, and 160 µm bands. All cutouts are centered on the \textit{Forges}.}
    \label{fig:tot_SED}
\end{figure}

\begin{figure}[H]
    \centering
    \begin{subfigure}{\linewidth}
        \centering
        \includegraphics[width=\linewidth]{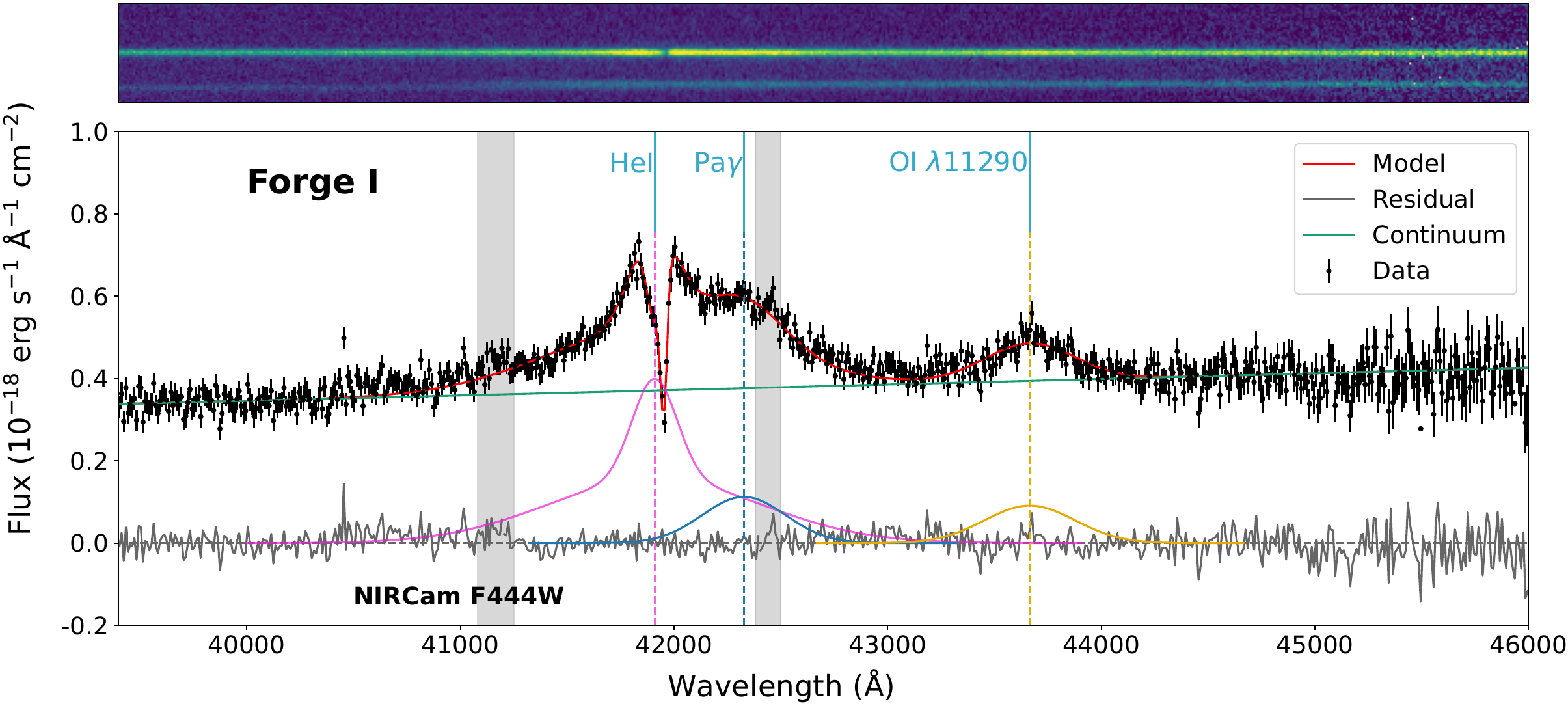}
    \end{subfigure}
    
    \vspace{10pt}
    
    \begin{subfigure}{\linewidth}
        \centering
        \includegraphics[width=\linewidth]{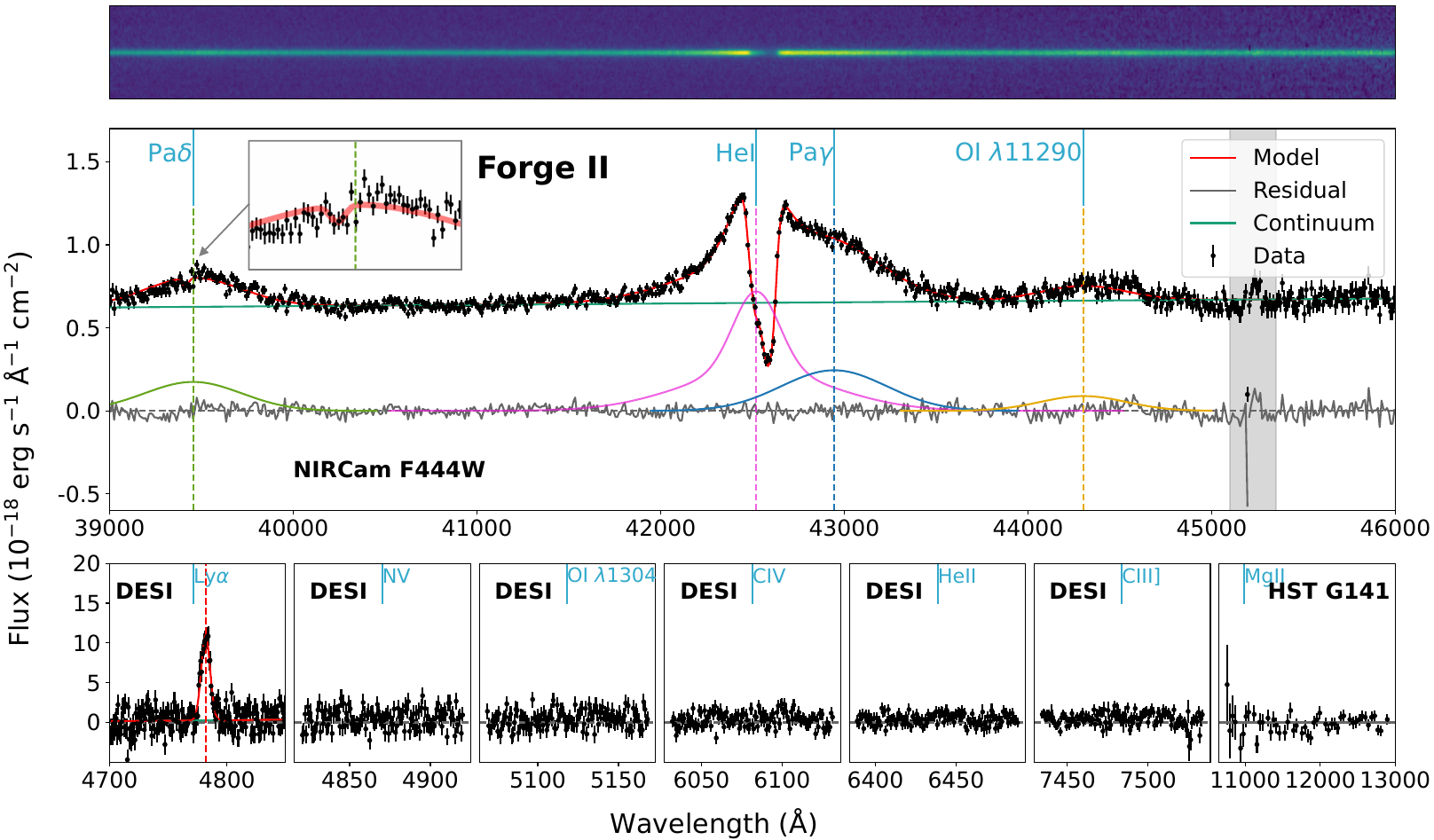}
    \end{subfigure}
    \caption{JWST NIRCam F444W slitless spectra of the \textit{Forges}. We also show the best-fit results and associated residuals. Prominent emission features (e.g., \ion{He}{i}, \ion{O}{i} $\lambda11287$, Pa$\gamma$, Pa$\delta$) are labeled. Short vertical lines at the top, labeled with line names, mark the theoretical positions of the emission lines. For \textit{Forge~II}, A zoom-in on the Pa$\delta$ absorption feature is shown in the subpanel, and additional spectroscopic coverage from DESI (covering the wavelength ranges of Ly$\alpha$, \ion{N}{v}, \ion{O}{i} $\lambda1304$, \ion{C}{iv}, \ion{He}{ii}, and \ion{C}{iii]}) and HST G141 (\ion{Mg}{ii}) is also displayed. The gray-shaded regions indicate the wavelength ranges masked during the fitting process due to contamination from neighboring sources in the NIRCam spectra.}
    \label{fig:linefit}
\end{figure}

In this work, we present two unique LRDs at $z = 2.868$ and 2.925. They show strong emission in the X-ray, radio, and mid-IR bands, offering a deep insight into the evolutionary path of LRDs. We hereafter refer to these two objects as \textit{Forge~I} and \textit{Forge~II}. The \textit{Forges} were found in the COSMOS field using the combination of JWST NIRCam (F090W, F115W, F200W, F277W, F356W, and F444W), HST (F814W), Subaru ($grizy$), and CFHT ($U$) images (Methods). They are unresolved in the JWST NIRCam F444W images, and their F444W magnitudes are 20.03 and 19.25 mag (magnitudes are in the AB system), respectively. They have blue UV slopes ($\beta_{\rm UV}=-1.23$ and $-0.73$) and red optical slopes ($\beta_{\rm opt}=2.91$ and 1.44) in the rest frame that satisfy the widely adopted LRD V-shaped SED criterion\cite{Kocevski2025} ($\beta_{\rm UV}<-0.37$ and $\beta_{\rm opt}>0$). The \textit{Forges} were also observed with JWST NIRCam F444W slitless spectroscopy. The spectra display broad emission lines of \ion{He}{i} $\lambda 10830$ (hereafter \ion{He}{i}), Pa$\gamma$, and \ion{O}{i} $\lambda 11287$ as well as \ion{He}{i} and Pa$\delta$ absorption lines that are frequently seen in the JWST spectra of LRDs\cite{2024MNRAS.535..853J,Lin2025_locallrd,Wang2025}. Figure~\ref{fig:tot_SED} shows the broadband SEDs and images of the \textit{Forges}, Figure~\ref{fig:linefit} shows the spectra, and Table~\ref{tab:bh_properties} lists the basic source information, photometry measurements and derived properties. We adopt a standard $\Lambda$CDM cosmology with $H_0 = 70\ \rm{km}\ \rm{s}^{-1}\ \rm{Mpc}^{-1}$, $\Omega_m = 0.3$ and $\Omega_\Lambda = 0.7$, where $H_0$ is the current value of the Hubble constant, and $\Omega_m$ and $\Omega_\Lambda$ are the cosmological density parameters for matter and dark energy, respectively.

\medskip

We estimate the central BH masses $M_{\mathrm{BH}}$ and bolometric luminosities $L_{\rm bol}$ of the \textit{Forges} using the spectral and photometric data above. We first model their continuum and line emission and absorption features in the JWST slitless spectra (Methods), and the best-fit results are listed in Extended Data Table~\ref{tab:linefits}. The full widths at half maximum (FWHM) of the emission lines are about several thousand km s$^{-1}$, confirming their AGN nature. In addition, the strong \ion{O}{i} $\lambda 11287$ emission agrees with the predicted signature of high-accretion SMBHs \cite{Inayoshi_2022}. Assuming that these lines originate from the AGN broad-line regions, we calculate BH mass using the broad Paschen emission lines and the 1 $\mu$m continuum luminosity\cite{Landt_2013}. The resultant masses are $(3.72 \pm 1.86) \times 10^8\,M_\odot$ and $(9.78 \pm 4.89) \times 10^8\, M_\odot$ ($M_{\odot}$ is the Solar mass), respectively. \textit{Forge~II} was also observed by the DESI spectroscopic survey and HST G141 slitless spectroscopy. Only narrow Ly$\alpha$ emission line (FWHM$=556 \rm ~km~\,s^{-1}$) is detected by DESI, shown in the lower panel of Figure~\ref{fig:linefit}. Canonical AGN broad lines, including Ly$\alpha$, \ion{N}{v} $\lambda1240$, \ion{C}{iv} $\lambda1549$, \ion{He}{ii} $\lambda1640$, and \ion{C}{iii]} $\lambda1909$, are not detected, which is consistent with previous results that high-redshift LRDs have weak or no broad UV emission lines  \citep{Greene2024,Wang2025}. We assume that the central AGNs are obscured in the rest-frame UV, but dominate in the near-IR, radio, and X-ray bands, then we obtain AGN bolometric luminosities $L_{\rm bol}$ by integrating their reprocessed SEDs (Extended Data Figure~\ref{fig:SED}). The inferred Eddington ratios are $L_{\rm bol}/L_{\rm Edd} = 0.17^{+0.25}_{-0.08}$ and $0.24^{+0.42}_{-0.12}$, where the Eddington luminosity is $L_{\rm{Edd}}=1.3\times 10^{38}\ (M_{\rm{BH}}/M_{\odot})\ \rm{erg\ s^{-1}}$.

\medskip

We also detect strong and slightly redshifted \ion{He}{i} absorption lines in the spectra of the two objects. The absorption indicates inflows or fallback \cite{Zhang_2017} that is different from commonly seen blueshifted troughs explained by AGN-driven outflows. Meanwhile, \textit{Forge II} has a blueshifted Pa$\delta$ absorption line that corresponds to an outflow velocity of $317.5$ km s$^{-1}$, implying complex gas kinematics. Such Pa$\delta$ absorption, like Balmer absorption, arises in dense and excited gas due to the short-lived populations of the $n=2$ or $n=3$ levels, and is much rarer than \ion{He}{i} absorption that has also been observed in some pre-JWST AGNs \cite{Zhang_2017,Hamann_2019}. Therefore, the detection of the Pa$\delta$ absorption further suggests the presence of high-density, low-ionization gas surrounding the \textit{Forges}. 

\medskip

While the lack of UV emission lines and the strong He I absorption make \textit{Forges} closely resemble LRDs, their X-ray detection reveals that they diverge in the level of obscuration and/or accretion activity. 
The \textit{Forges} have X-ray observations by the XMM-Newton Wide-Field Survey\cite{2007ApJS..172...29H} and the Chandra COSMOS Legacy Survey\cite{2016ApJ...819...62C}. Both sources are X-ray luminous, with observed 2--10 keV X-ray luminosity $L_{\mathrm{X}}=2.6\times10^{44}\ \rm{erg\ s}^{-1}$ and $8.9\times10^{44}\ \rm{erg\ s}^{-1}$, respectively (Methods). The best-fit gas column densities ($N_{\rm H}$) are $2.1^{+1.8}_{-1.3} \times 10^{22} \,\rm{cm}^{-2}$ and $1.4^{+1.8}_{-1.3} \times 10^{22}\, \rm{cm}^{-2}$. The modest X-ray obscuration, along with the non-detection of broad UV lines, indicates that the \textit{Forges} are different from heavily obscured AGNs.
Their bolometric-to-X-ray luminosity ratios, $L_{\mathrm{bol}} / L_{\mathrm{X}} \simeq 43$ and $\simeq 51$, are smaller than those of most high-redshift LRDs\cite{Inayoshi2024_xrayweak}. For \textit{Forge II}, we also detect the 6.4 keV Fe K$\alpha$ line with a confidence level of $98.4\%$ ($p = 0.016$) (Extended Data Figure~\ref{fig:Xray_spec}), which is a characteristic X-ray feature of AGNs. Both sources exhibit strong X-ray variability (Extended Data Figure~\ref{fig:Xray_lc}). The $L_{\mathrm{X}}$ variability of \textit{Forge~I} reaches a maximum amplitude of $5.4^{+37.8}_{-2.7}$ (defined as the ratio of the maximum to minimum flux) over a timescale of $\sim 2$ years in the observed frame. \textit{Forge~II} shows a variability amplitude of $1.7^{+0.8}_{-1.0}$ over $\sim 0.5$ years. In both cases, the observed variability exceeds what is typically seen in AGNs\citep{Timlin2020}. In contrast to their X-ray variability, they show no apparent variability in the F115W-band images over a baseline of two years. These behaviors are consistent with the prediction for high-accretion LRDs, where photon trapping suppresses UV variability while allowing strong X-ray fluctuations\citep{Inayoshi2024_xrayweak}. 
Our following analysis indicates that the measured UV flux is dominated by extended emission rather than by the central AGNs (Extended Data Figure~\ref{fig:galfitm_ID1} and \ref{fig:galfitm_ID2}), which further dilutes the variability amplitude.  
The observed UV–to–X-ray power-law spectral slope $\alpha_{\rm ox}=f_{2\rm \ keV}/f_{2500\ \text{\AA}}$ largely deviates from the $\alpha_{\rm ox}-L_{2500\ \text{\AA}}$ relation for AGNs, where $f_{2500\ \text{\AA}}$ is the measured rest-frame 2500~\AA\ flux dominated by the host galaxy. 
Using the intrinsic $f_{2500\ \text{\AA}}$ inferred from the ionized nebular emission (see below and Extended Data Figure~\ref{fig:Nebular_fit}), the corrected $\alpha_{\rm ox}$ agrees with the observed X-ray luminosity (Extended Data Figure \ref{fig:aox_FP}). These results imply that the intrinsic UV radiation is generally consistent with normal AGNs.

\medskip

The \textit{Forges} also exhibit radio emission, a property commonly seen in normal AGNs.
They are both detected by the VLA-COSMOS deep 3 GHz survey \cite{Smolcic_2017}, with flux densities of $15.2 \pm 2.4\,\mu$Jy and $83.9 \pm 4.8\,\mu$Jy, respectively. Furthermore, \textit{Forge~II} is detected by the VLA-COSMOS Large Project \citep{Schinnerer_2004_VLAlarge} with a flux density of $183 \pm 25\,\mu$Jy at 1.4 GHz. The derived radio spectral index is $\alpha = -1.02 \pm 0.19$ (where $f_\nu \propto \nu^\alpha$), consistent with the distribution of spectral indices ($\alpha = -0.73\pm0.35$) in large radio AGN samples\citep{Smolcic_2017}. Adopting this spectral index for both sources yields observed or apparent radio-loudness parameters of $R = 28.3$ and 48.5, where $R$ is defined as $R = f_{\rm 5GHz}/f_{4400\ \text{\AA}}$ and $f_{4400\ \text{\AA}}$ is the rest-frame 4400~\AA\ flux derived from the F115W- and F200W-band flux. Using the intrinsic $f_{4400\ \text{\AA}}$ inferred from the ionized nebular model below (Extended Data Figure~\ref{fig:Nebular_fit}), the radio loudness values decrease to $R = 5.2$ and 1.5 that are both below the conventional threshold for radio-loud AGNs ($R > 10$). We further find that the \textit{Forges} are broadly consistent with radio-quiet AGNs\citep{2024A&A...689A.327W} in the fundamental plane (Extended Data Figure \ref{fig:aox_FP}). This suggests that their X-ray and radio emission may already reach the level of normal AGNs, but the UV emission is still largely suppressed.

\medskip

We perform multi-band image decomposition to understand the different components of the \textit{Forges} (Methods). Similar to other LRDs, the \textit{Forges} become increasingly point-source dominated from $\lambda_{\rm obs} \sim 2 \,\mu \rm m$ ($\lambda_{\rm rest} \sim 5000$ \AA) toward longer wavelengths, while appearing extended at shorter wavelengths in the JWST images. Our initial decomposition procedure uses a model with one point spread function (PSF) and one Sérsic component, and reveals significant off-center residuals that are particularly strong in the F200W and F277W bands. These two bands cover [\ion{O}{iii}]+H$\beta$ and H$\alpha$, respectively. Therefore, 
we add an additional Sérsic profile to represent the off-center component in the model, and we attribute this component as nebular emission rather than stellar emission\citep{Chen2025b}, as supported by its morphology and large line equivalent width (EW). Extended Data Figures~\ref{fig:galfitm_ID1} and \ref{fig:galfitm_ID2} and Extended Data Table~\ref{tab:galfitm_res_both} present our image decomposition results. These results disclose an emergence of point-source emission that is weak in the UV but dominates the optical to near-IR radiation. This is consistent with the non-detection of high-ionization UV emission lines from AGNs.

\medskip
With the image decomposition results above, we model the SEDs of the \textit{Forges} for different components, including the central AGN, host galaxy, and off-center emission (Methods). The best-model results show that the optical continuum of the central point source is well reproduced by a high-accretion SMBH surrounded by an optically thick photosphere with effective temperature $T_{\mathrm{eff}}\sim5000$~K\cite{Kido2025,LiuH_2025}. The $\sim 1$--$6\ \mu$m emission excess requires an additional component with $T_{\mathrm{eff}}\sim500$~K that plausibly traces radiation from an emerging hot-dust torus. There is evidence that lower-redshift LRDs tend to exhibit enhanced IR emission compared to their high-redshift counterparts\cite{2025arXiv250907100D}, 
pointing to the onset of dust-torus formation. The \textit{Forges} appear to represent a more evolved stage, where the dusty structure is substantially more developed.
Attributing all the extended emission to stars would imply unrealistically high stellar masses ($>10^{12}\,M_\odot$), suggesting a significant contribution from non-stellar light. The off-center emission is strong in the F200W and F277W bands, corresponding to [\ion{O}{iii}]/H$\beta$ and H$\alpha$, indicating photoionized gas\citep{Chen2025b}. A stellar ionized nebula cannot reproduce such strong off-center emission as revealed by the SED fitting.
We therefore consider only the central extended component (Sérsic 1) as the host galaxy, while attributing the remaining off-center extended emission to gas clouds photoionized by the central AGN\cite{Chen2025b}.
The host galaxies have moderate star formation rates (SFRs) of $16.0 \pm 0.9$ and $77.7 \pm 6.6\ M_\odot\,\rm{yr}^{-1}$ and stellar masses of $\log (M_\ast/M_\odot)=9.40 \pm 0.18$ and  $10.28 \pm 0.16$, consistent with the star-forming main sequence at $z\sim3$ \cite{Clarke_2024}. They dominate the UV SEDs of the two objects. This is further supported by a pronounced 2175~\AA\ absorption feature that represents the presence of graphitic dust grains\cite{2021ApJ...909..213K}. 
For the off-center extended emission, we use \texttt{CLOUDY} to model the observed SEDs as nebular gas illuminated by the central AGN through an opening angle, exploring a grid of intrinsic AGN luminosities and nebular properties. 
The best-model intrinsic AGN UV luminosities are substantially higher than the observed UV luminosities of the point sources.
In this scenario, the bulk of the UV emission from the center is absorbed by the optically thick photosphere and re-emitted as blackbody radiation with $T_{\mathrm{eff}}\sim5000$~K\cite{Kido2025,LiuH_2025}. Part of ionizing photons directly escape along the polar direction that is perpendicular to the line of sight, and photoionize the surrounding nebula\cite{2025arXiv250506359T}. The intrinsic AGN UV luminosity inferred from the \texttt{CLOUDY} modeling of the nebular emission agrees with the observed optical–IR integrated luminosity (Extended Data Figure \ref{fig:Nebular_fit}), validating an overall energy balance between the absorbed and reprocessed emission. 

\medskip

\begin{figure}[H]
    \centering
    \includegraphics[width=\linewidth]{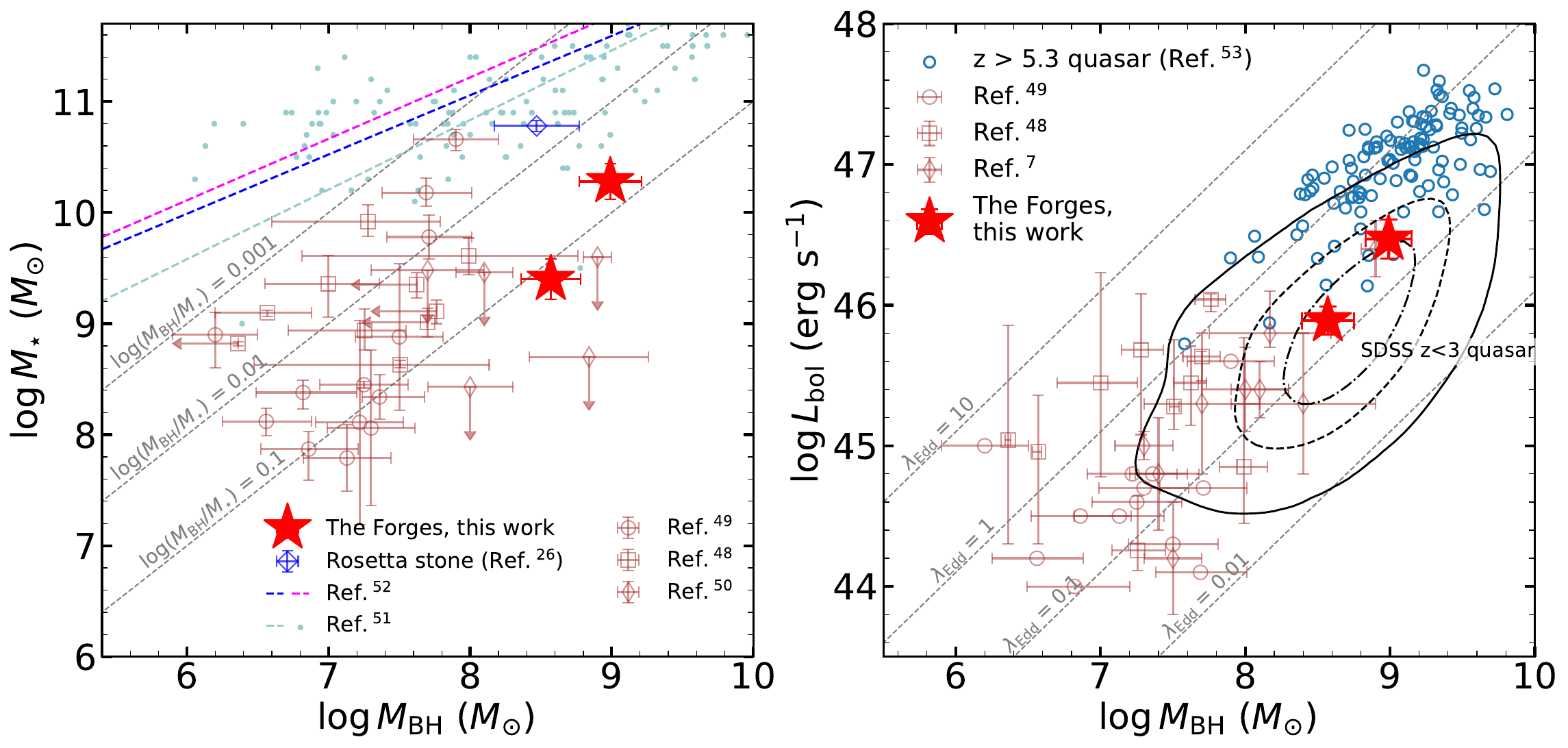}
    \caption{\textbf{Left:} $M_*$ versus $M_{\mathrm{BH}}$ for the \textit{Forges}, shown as red stars. For comparison, we include a few high-redshift LRD and AGN samples (brown symbols)\cite{2023ApJ...959...39H,Maiolino2024_diverse, 2025ApJ...983...60C}, a $z=2.26$ LRD (blue dot)\cite{2024MNRAS.535..853J}, as well as local scaling relations: inactive galaxies (light blue dots and dashed line)\cite{2020ARA&A..58..257G}, AGN with early-type hosts (blue dashed line), and AGNs with late-type hosts (pink dashed line)\cite{2023NatAs...7.1376Z}. The grey dashed lines denote constant $M_{\mathrm{BH}}$–to–$M_*$ ratios. 
    \textbf{Right:} $L_{\mathrm{bol}}$ versus $M_{\mathrm{BH}}$. For comparison, we include high-redshift quasar (blue open circles)\cite{ 2023ARA&A..61..373F} and LRD/AGN samples (brown symbols)\cite{2023ApJ...959...39H, Maiolino2024_diverse, Greene2024}. The SDSS quasars with z $\leq$3 are shown in the black contour\cite{2025ApJ...988..204C}. The grey dashed lines indicate constant Eddington ratios. }
    \label{fig:MBH_Ms}
\end{figure}

\begin{figure}[H]
    \centering
    \includegraphics[width=0.93\linewidth]{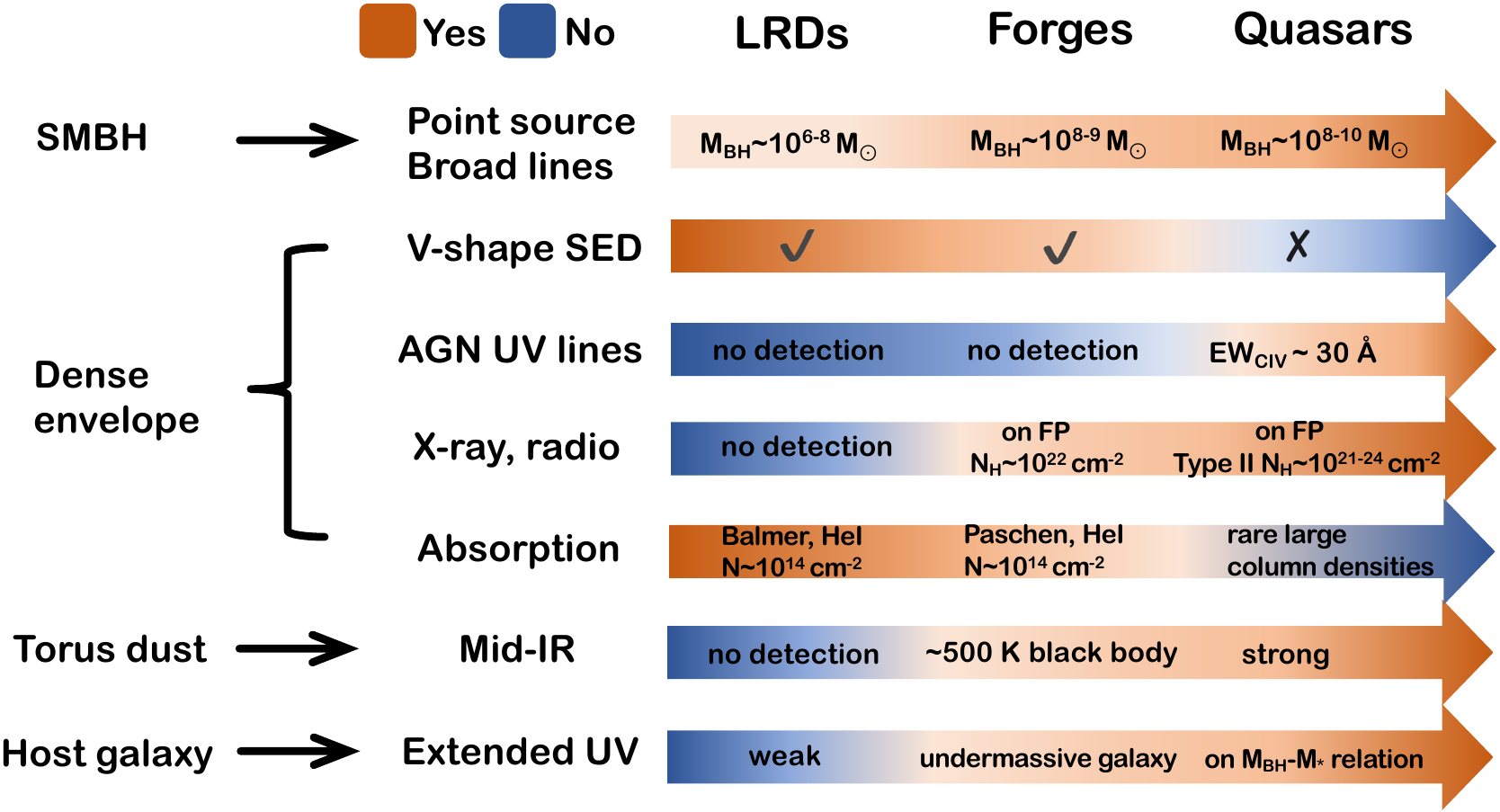}
    \caption{ Comparison between LRDs, the \textit{Forges} and typical quasars. The \textit{Forges} occupy a transitional stage between LRDs and quasars.}
    \label{fig:compare}
\end{figure}

LRDs are often thought to be super-Eddington accreting SMBHs enshrouded in dense gas ($n_{\rm H} \sim 10^{9}$--$10^{11}\,\rm cm^{-3}$)\cite{Maiolino_JWSTChandra,2025ApJ...980L..27I,Lin2025_locallrd}, and their blue UV emission possibly arises from stellar population \cite{2024ApJ...969L..13W,Kocevski2025,2025arXiv250919422I} or photoionized nebular clouds\cite{Chen2025b}. The \textit{Forges} can be explained by this scenario in the context of evolution. The dense gas envelope around LRDs is expected to disperse and the LRD phase terminates due to feedback or accretion onto the central BH when $M_{\rm BH}$ becomes sufficiently large and the envelope structure can no longer be maintained\citep[][]{2025arXiv250919422I,Kido2025}. The \textit{Forges} are witnessing this transitional phase during which LRDs are evolving into quasars. As shown in Figure~\ref{fig:MBH_Ms}, the SMBHs are overmassive compared with the local $M_{\rm BH}$--$M_*$ relation, while their host galaxies are already well assembled and lie on the star-forming main sequence at $z\sim3$. On the $L_{\rm bol}$--$M_{\rm BH}$ plane, the \textit{Forges} start to deviate from the LRD population and share comparable $L_{\rm bol}$ and $M_{\rm BH}$ with lower-redshift quasars. LRDs at high redshift ($z\gtrsim4$) are generally X-ray-weak, as they are embedded within the dense gas envelope that suppresses the escape of X-ray photons\cite{Maiolino_JWSTChandra,Yue2024,Inayoshi2024_xrayweak,LiuH_2025,Kido2025}. In X-ray detected LRDs (e.g., Ref.\cite{Kocevski2025}), including the \textit{Forges}, the gas envelope is likely dispersing into a clumpy and porous structure. The emerging X-ray emission can escape along the line of sight, while most UV photons remain trapped (or dust reddened) owing to a much smaller X-ray cross-section and/or a smaller X-ray emission region. Radio jets are weakly confined, giving rise to the observed radio emission. Along the polar directions, the gas envelope is being cleared rapidly by AGN outflows, so UV ionizing photons escape easily through an opening angle and photoionize the surrounding gas\cite{2025arXiv250506359T}. Meanwhile, a dust torus starts to form and dominate the mid-IR SED. At this evolutionary stage, different viewing angles yield distinct observational properties. With this scenario, the \textit{Forges} will eventually evolve into normal quasars. 
The evolutionary pathway and key observational properties from LRDs to quasars are summarized in Figure~\ref{fig:compare}.
From the contiguous slitless spectroscopy of COSMOS-3D, we identify seven broad \ion{He}{i} LRDs at $z\sim3$ (see Methods), two of which show the transitional properties as presented here. This implies that roughly 30\% of LRDs at this epoch are observed in this short-lived stage, while the fate of the remaining LRDs remains unclear.
Our finding provides a key link between some LRDs and normal AGNs.

\newpage
\section*{Figures and Tables}

\begin{table}[H]
\centering
\caption{Physical properties of the \textit{Forges}.}
\label{tab:bh_properties}
\begin{tabular}{lcc}
\hline
Property & \textit{Forge~I} & \textit{Forge~II} \\
\hline
RA & 09:59:30.28 & 09:59:34.07   \\  
DEC & 02:10:00.12  & 02:17:06.31\\       
Redshift & 2.868 & 2.925 \\
\hline
CFHT $U$ & $27.26 \pm 0.07$ & $26.37 \pm 0.06$ \\
Subaru $g$ & $26.16 \pm 0.07$ & $25.00 \pm 0.06$ \\
Subaru $r$ & $25.58 \pm 0.06$ & $24.41 \pm 0.06$ \\
Subaru $i$ & $25.59 \pm 0.06$ & $24.20 \pm 0.05$ \\
Subaru $z$ & $25.74 \pm 0.07$ & $24.30 \pm 0.07$ \\
Subaru $y$ & $25.75 \pm 0.07$ & $24.13 \pm 0.06$ \\
HST F814W & $25.74 \pm 0.05$ & $24.30 \pm 0.05$ \\
JWST F115W & $25.12 \pm 0.09$ & $23.76 \pm 0.06$ \\
JWST F150W & $24.31 \pm 0.07$ & $22.88 \pm 0.06$ \\
JWST F200W & $22.78 \pm 0.05$ & $21.81 \pm 0.05$ \\
JWST F277W & $21.27 \pm 0.05$ & $20.48 \pm 0.05$ \\
JWST F356W & $20.81 \pm 0.05$ & $19.99 \pm 0.05$ \\
JWST F444W & $20.03 \pm 0.05$ & $19.25 \pm 0.05$ \\
JWST F770W & $18.52 \pm 0.12$ & $17.76 \pm 0.12$ \\
JWST F1280W & $18.04 \pm 0.12$ & $17.15 \pm 0.12$ \\
JWST F2100W & $17.21 \pm 0.12$ & $16.07 \pm 0.12$ \\
Spizter MIPS24 & $17.33 \pm 0.12$ & $15.93 \pm 0.12$ \\
Spizter MIPS70 & $<16.58^{a} $ & $<16.07$ \\
Spizter MIPS160 & $<14.46 $ & $<14.23$ \\
ALMA 1.1 mm ($\mu$Jy) & \textemdash & $<30.08$ \\
\hline
$\beta_{\rm UV}$ &  $-1.23\pm0.06$ & $-0.73\pm0.02$\\
$\beta_{\rm optical}$ & $2.91\pm0.12$ & $1.44\pm0.06$\\
$\Gamma_{\rm 0.3-10keV}$ & $1.80^{+0.19}_{-0.18}$ & $1.54\pm0.09$\\
$\alpha_{r}$ & \textemdash & $-1.02\pm0.19$\\
\hline
$\log [M_{\rm BH}(M_\odot)]$ & $8.57 \pm 0.21$ & $8.99 \pm 0.22$ \\
$\log [L_{\rm bol}(\rm{erg\ s}^{-1})]$ & $45.89 \pm 0.1$ & $46.47 \pm 0.14$ \\
$\lambda_{\rm Edd}$ & $0.17_{-0.08}^{+0.25}$ & $0.24_{-0.12}^{+0.42}$ \\
$\log [L_{\rm X}\,(\rm erg\,s^{-1})]$ & $44.41 \pm 0.11$ & $44.95 \pm 0.14$ \\
\hline
$\log [M_\ast(M_\odot)]$ & $9.40 \pm 0.18$ & $10.28 \pm 0.16$ \\
SFR (M$_\odot$\,yr$^{-1}$) & $16.0 \pm 0.9$ & $77.7 \pm 6.6$ \\
$\log [Z_\ast(Z_\odot)]$ & $-1.70 \pm 0.15$ & $-1.28 \pm 0.60$ \\
Age (Gyr) & $0.43 \pm 0.33$ & $0.69 \pm 0.38$ \\
$A_V$ (mag) & $1.36 \pm 0.13$ & $1.36 \pm 0.30$ \\
$N_{\rm H} \, (\rm{cm}^{-2})$ &  $2.1^{+1.8}_{-1.3} \times 10^{22}$ & $1.4^{+1.8}_{-1.3} \times 10^{22}$ \\
\hline
\end{tabular}
\begin{flushleft}
\textbf{Notes.} \\
$^a$~ 2$\sigma$ Upper limit.
\end{flushleft}
\end{table}  

\newpage
\section*{Methods}

\medskip
\noindent {\bf Imaging and photometric data} 

{\bf JWST NIRCam and HST data}. The \textit{Forges} were observed by two JWST imaging programs: COSMOS-Web (GO\#1727\cite{Cweb_survey_ref}; PIs: J. Kartaltepe and C. Casey) and COSMOS-3D (GO\#5893; PI: K. Kakiichi) using the near-IR Camera \cite{NIRCAM_ref} with filter bands of F115W, F150W, F200W, F277W, F356W, and F444W. We reduced the NIRCam imaging data following the steps in Ref.\cite{2025ApJ...985..119Z}. Briefly, we reduced the raw data using the combination of the \textit{JWST Calibration Pipeline} (v.1.17.0) with CRDS version 11.17.0 and CRDS context 1185, the CEERS\cite{2023ApJ...946L..12B} NIRCam imaging reduction \footnote{\url{https://github.com/ceers/ceers-nircam.}} and our own custom codes. All images were aligned to the HST images from the HST COSMOS survey\footnote{\url{https://irsa.ipac.caltech.edu/data/COSMOS/images/acs_mosaic_2.0/}.}. These HST images were also included in our analyses below. The final NIRCam and HST images  have a pixel scale of $0\farcs03$. 

{\bf JWST MIRI data}. The MIRI images of the two LRDs were observed by COSMOS-Web (F770W) and COSMOS-3D (F1000W and F2100W). We reduced the MIRI data using the \textit{JWST Calibration Pipeline} (v.1.17.0) with CRDS version 11.17.0 and CRDS context 1185. Our reduction pipeline is based on the public CEERS pipeline \cite{2023ApJ...956L..12Y},\footnote{\url{https://github.com/ceers/ceers-miri}.} with several custom steps to improve data quality. The details can be found in Ref.\cite{2025ApJ...985..119Z} In addition to the standard Stages 1 and 2 processing, we removed the horizontal and vertical stripes after Stage 2 and mitigated warm-pixel contamination by masking pixels identified in median-stacked images. We also adopted a modified background-subtraction scheme \cite{2024ApJ...976..224A}, in which a super-background was constructed from other exposures with sources masked, followed by a large-scale filtering. All images were aligned to the NIRCam F444W mosaic image. The final MIRI images have a pixel scale of $0\farcs06$. 

{\bf Subaru, CFHT, and Spitzer data}. We used Ultradeep optical ($grizy$) imaging data from the Hyper Suprime-Cam Subaru Strategic Program \cite{2022PASJ...74..247A} third public data release (PDR3). We incorporated the CFHT $U$-band images from the MUSUBI (MegaCam Ultra-deep Survey: u*-band Imaging) survey \cite{2022ApJS..260...54W}. At longer wavelengths, we included Spitzer MIPS 24, MIPS 70, and MIPS 160 images from the S-COSMOS survey \citep{2007ApJS..172...86S}.

For the JWST NIRCam and HST images, we constructed empirical PSF models for each band using \texttt{PSFEx} \citep{2011ASPC..442..435B}, following the method in Ref.\citep{2024ApJ...962..139Z}. We used these PSFs to derive kernels to match all PSFs to the F444W PSF using \texttt{PyPHER} \citep{2016ascl.soft09022B}, and then used them to convolve all images to match the F444W image. Next, we performed photometry using \texttt{SExtractor} \citep{1996A&AS..117..393B} on the PSF-matched images to construct source catalog. We used the dual image mode and the detection image is a $\chi^2$ image combining the PSF-matched NIRCam images. Photometry is measured in a small Kron aperture ($k$=1.1, $R_{\min}$=1.6), and an aperture correction is performed to match the default Kron aperture ($k$=2.5, $R_{\min}$=3.5) in each band. The photometric errors is measured following the method in Ref.\cite{Zhang2025b} by placing random apertures. 

For the HSC and CFHT images, we derived template-fitting deblending photometry to get the total fluxes at each band with T-PHOT v2.0\citep{2016A&A...595A..97M}. T-PHOT
uses position priors from the high-resolution NIRCam F444W images. We generated PSF models for the NIRCam, HSC, and CFHT images using \texttt{PSFEx}, and constructed convolution kernels to match the NIRCam F444W PSF to those of the HSC and CFHT images. These kernels were used to build image templates for each source, from which the photometry in each band is extracted through template fitting with T-PHOT.

The MIRI and Spitzer images have much larger PSFs than those of the other bands, so the sources are unresolved. We performed aperture photometry and applied aperture corrections separately. For the MIRI data, we adopted an aperture radius corresponding to 65\% of the encircled energy as given in Ref.\citep{2024ApJ...976..224A}, followed by an aperture correction. The photometric uncertainties were estimated from the background fluctuations within an annulus larger than the PSF size with other sources masked. For the Spitzer images, we applied a similar procedure using the recommended aperture radius, aperture correction, and background annulus\footnote{\url{https://cosmos.astro.caltech.edu/page/spitzer}.}.

{\bf VLA data}. The two sources were detected in the VLA-COSMOS deep 3~GHz survey \citep{Smolcic_2017}.The synthesized beam size is $0\farcs75$ (circularized) across the two-square-degree field. The local rms at the source positions is $2.3\ \mu{\rm Jy\ beam^{-1}}$ for \textit{Forge~I} and $2.23\ \mu{\rm Jy\ beam^{-1}}$ for \textit{Forge~II}. \textit{Forge~I} has a flux density of $15.2 \pm 2.4\ \mu$Jy, and was classified as \textit{unresolved} in the catalog. \textit{Forge~II} has a flux density of $83.9 \pm 4.8\,\mu$Jy, and is \textit{resolved} with a size over several synthesized beams. 

In the 1.4~GHz VLA-COSMOS Large Project catalog \citep{Schinnerer_2004_VLAlarge}, \textit{Forge~II} (COSMOSVLA\_J095934.07+021706.4) has an integrated flux density of $183 \pm 25\,\mu$Jy and a deconvolved size of $1\farcs10\times0\farcs41$ at a position angle of $124^\circ$, corresponding to a linear size of $\sim9$~kpc. The rms at the radio source position is $0.178\,{\rm\ mJy\ beam^{-1}}$. The synthesized beam size is $1\farcs5\times1\farcs4$. 

From the 1.4–3~GHz flux ratio, we derived a radio spectral index of $\alpha = -1.02 \pm 0.19$ (where $f_\nu \propto \nu^\alpha$), consistent with the distribution of spectral indices in large radio AGN samples (mean $\alpha = -0.73$, $\sigma = 0.35$\cite{Smolcic_2017}). 
Adopting this spectral index for both sources yields radio-loudness parameters of $R = 28.3$ and $R = 48.5$, where $R = f_{5\,\mathrm{GHz}}/f_{4400\,\text{\AA}}$ \citep{Kellermann_1989} and $f_{4400\,\text{\AA}}$ is derived from the F115W and F200W-band flux. 
As described in the SED modeling section of the Methods, a large fraction of the $f_{4400\,\text{\AA}}$ flux from the accretion disk of the central AGN is absorbed along the line of sight and re-emitted in the rest-frame optical and near-IR. Consequently, these $R$ values are overestimated. If we instead adopt the intrinsic $f_{4400\ \text{\AA}}$ inferred from the ionized nebular model (Extended Data Figure~\ref{fig:Nebular_fit}), the radio loudness values decrease to $R = 5.2$ and 1.5 that are both below the conventional threshold for radio-loud AGNs ($R > 10$).

{\bf ALMA data.}
\textit{Forge~II} was covered by an archival ALMA program (PID 2024.1.01085.S). The observation was made on band 6 with an integration time of 890 s. We used the publicly available ALMA continuum images from the $\textrm{A}^3$COSMOS project\cite{Liu_2019_A3cosmos}, which systematically reprocessed all archival ALMA observations in the two-square-degree COSMOS field to produce homogeneous, high-sensitivity (sub)millimeter maps.
The $\textrm{A}^3$COSMOS ALMA continuum images were produced with the CASA pipeline\cite{2007ASPC..376..127M} (v4.7.2) in ``continuum'' + ``automatic'' mode with Briggs weighting (robust = 2) and multi-frequency synthesis (nterms = 2). These maps achieve on average $\sim10\%$ better rms sensitivity than the official ALMA QA2 products, with typical angular resolutions characterized by a synthesized beam major-axis FWHM of $\sim1''$. \textit{Forge~II} was not detected in the ALMA image. We estimated the rms value at the position of \textit{Forge~II} and corrected the PB value, resulting in a $2\sigma$ upper limit of $30 \mu\rm Jy$.

\medskip
\noindent {\bf Spectroscopic data} 

{\bf JWST NIRCam slitless spectra.} COSMOS-3D (GO5893, PI: K.~Kakiichi) obtained NIRCam WFSS observations of the COSMOS field using the row-direction grism (Grism R) combined with the wide-band filter F444W. The WFSS data were reduced following the standard procedure \cite{Sun_2023}, and the corresponding codes and calibration files are publicly available\footnote{\url{https://github.com/fengwusun/nircam_grism}}.
We first processed the NIRCam WFSS data using the standard JWST Stage-1 calibration pipeline (version \verb|1.13.4|) and the calibration reference file \verb|jwst_1364.pmap|. A flat-field correction was applied using flat-field data obtained with the same filter and detector. For each exposure, we constructed and subtracted a super-sky background using all flat-fielded exposures taken with the same filter+pupil+module combination. Residual background structure remained after this step, so we performed an additional 2D sky-background subtraction using the Source Extractor algorithm \cite{1996A&AS..117..393B}. The background was estimated on the $2.5\sigma$-clipped image using a mode estimator of the form $(2.5 \times \mathrm{median}) - (1.5 \times \mathrm{mean})$, which is less sensitive to source crowding than a simple clipped mean. If $(\mathrm{mean} - \mathrm{median}) / \mathrm{STD} > 0.3$ (where $\mathrm{STD}$ denotes the standard deviation), the mode estimator becomes unreliable and a simple median is used instead. Finally, the astrometry of the grism data was refined by matching the simultaneously obtained short-wavelength images to the long-wavelength source catalog, which had previously been registered to the HST catalog.

{\bf HST G141 and DESI spectra.} 3D-DASH (GO16259/16443, PI: Ivelina Momcheva) observed the entire 1.7 deg$^2$ COSMOS field with the broadband images in WFC3 F160W and the parallel slitless spectroscopy in WFC3 G141. It covered \textit{Forge~II} with both direct images and slitless spectroscopy. Each mode took 1774 s of exposure time. We reduced the spectrum using the pipeline GRIZLI\footnote{\url{https://grizli.readthedocs.io/en/latest/}}, similar to the procedure in previous works\cite{2025RAA....25b5015C}. We first downloaded the data from MAST. The raw exposures were bias- and dark-subtracted and flat-fielded. Preprocessing steps included calibrating the WCS of each direct+grism image pair, coadding the drizzled images, aligning the grism exposures when multiple position angles were available, and subtracting the background. A photometric catalog was constructed from the direct images for subsequent extraction. For each detected source, an initial flat continuum scaled to its broadband flux was assigned; for bright sources, the model was iteratively refined to higher-order polynomials by direct spectral fitting. This process produced the contamination model, which was subtracted from each grism image. The 2D beam of our target sources were then extracted with forward modeling: for all beams overlapping the targets, we ranked them by the F160W luminosity. Starting from the brightest, we iteratively extracted their 2D beams, fitted continuum models, and subtracted them from the target beam. Finally, using the 2D morphology of the source in F160W, we defined the extraction aperture and derived the final 1D spectrum. 
\textit{Forge II} is also covered by DESI spectroscopic observations, and we obtained its DESI spectrum from the archive\citep{2025arXiv250314745D}.

\medskip
\noindent {\bf Spectral fitting} 

We model the F444W slitless spectra of both sources using a power-law continuum, multiple Gaussian components for the emission lines, and a standard attenuation model for the \ion{He}{i} and Pa$\delta$ absorption features. The emission line set includes \ion{He}{i} (broad and narrow components), Pa$\gamma$, Pa$\delta$, and \ion{O}{i} $\lambda11287$.
Assuming the source is fully covered by the absorber, the absorption features are modeled using\cite{Savage1991} 
\begin{equation}
    f_\lambda=e^{-\tau_\lambda},
\end{equation}
where $\tau_\lambda$ is the optical depth profile
\begin{equation}
    \tau_\lambda = \tau_0\exp{\left[ -0.5\left(\frac{\lambda-\lambda_0e^{\Delta v/c}}{\sigma}\right)^2 \right]},
\end{equation}
in which $\tau_0$ is the optical depth at the core of the line, $\Delta v$ is the velocity shift of the absorber and $\sigma$ is its velocity dispersion. For \ion{He}{i}, the absorption trough shows an asymmetric profile, therefore two absorption components are required. For \textit{Forge~II}, Pa$\delta$ also shows a blueshifted absorption feature, which is fitted with one absorption model. All emission line components are fitted with their central wavelengths tied to a common systemic redshift parameter. For \textit{Forge~II}, we additionally link the line widths of Pa$\gamma$ and Pa$\delta$ to reduce the degeneracy due to blending and low signal-to-noise ratio (S/N). We perform a Markov chain Monte Carlo likelihood fit using \texttt{emcee} to constrain the model parameters. The wavelength ranges in the NIRCam slitless spectra that are contaminated by emission lines from neighboring sources are masked during the fitting.

The fitting result is shown in Figure~\ref{fig:linefit} and the key emission and absorption line properties  are summarized in Extended Data Table~\ref{tab:linefits}. The detection of deep and slightly redshifted \ion{He}{i} absorption troughs suggests inflowing material or fallback motions, rather than the commonly observed large blueshifted outflows in many AGNs. Meanwhile, the marginally detected blueshifted Pa$\delta$ absorption further indicates a complex multi-component absorbing system. The column density of the absorbers are computed as\cite{Savage1991} 
\begin{equation}
N=\frac{m_ec}{\pi e^2 f_0 \lambda_0}\tau,
\end{equation}
where $\tau$ is the integrated line optical depth and $f_0 \lambda_0$ is the product of the oscillator strength of the transition and the rest-frame wavelength. The calculated values of $N_{\rm He}(2^3\rm S)$ are shown in Table \ref{tab:linefits}, which are in the order of $\sim 10^{13.5}\rm \,cm^{-2}$. The estimated $N_{\rm He}(2^3\rm S)$ values are consistent with those reported in other LRDs \cite{2024MNRAS.535..853J,Wang2025,Federica2025}.
A detailed analysis of the strong \ion{He}{i} absorption system in the \textit{Forge II} is also presented in Z. Li et al. (submitted), who report \textit{Forge II} and another object as obscured X-ray AGNs. They further calculate the density of the absorbing gas of \textit{Forge II} as $\log (n_{\rm H}/\rm cm^3) = 10 $, with a distance of $\sim 0.02\,\rm pc$ to the ionizing source. These density and distance is comparable the expectation of the dense envelope of LRDs\citep{2025ApJ...980L..27I,Kido2025,Lin2025_locallrd}, suggesting that the absorbers may be related to the envelope.
\ion{He}{i} absorption probes partially ionized, high-column density gas with a large amount of helium pumped to the long-lived metastable triplet state $2^3S$. In contract, Pa$\delta$ absorption requires a significant population of hydrogen in the short-lived $n=3$ level. Like Balmer absorption, it forms only in high-density excited gas, much less frequently than \ion{He}{i} absorption\cite{2024MNRAS.535..853J,Wang2025}. 
Future JWST/NIRSpec MSA G395H observations will provide deeper and contamination-free spectra than the existing NIRCam WFSS data and enable a more detailed investigation of the absorbers and the properties of the ionizing radiation field.

The black hole masses estimated from Pa$\gamma$ and Pa$\delta$ \cite{Landt_2013} are $M_{\rm BH} = (3.70 \pm 1.85) \times 10^8~\rm M_\odot$ for \textit{Forge I} and $(9.75 \pm 4.88) \times 10^8~\rm M_\odot$ for \textit{Forge~II}, with additional 50 percent uncertainty included to account for the intrinsic scatter of the calibration\cite{Landt_2013}. These estimates assume that the lines are Doppler-broadened by the virial motion of gas in the broad-line region (BLR), and the BLR size is calibrated using local Type I AGNs. It remains uncertain whether the gas is truly virialized and whether LRDs follow the same radius–luminosity relation as Type I AGNs. Such assumptions may introduce systematic uncertainties of up to one order of magnitude\cite{LiuH_2025, Lupi_2024,2025arXiv250905434G}. 

According to some theoretical models for growing seed BHs\cite{Inayoshi_2022}, the \ion{O}{i} $\lambda11287$ line arises from Ly$\beta$ fluorescence pumping followed by a cascade\cite{Rodriguez_2002_OI}, tracing a high-density partially neutral zone in the outer disk or the inner broad-line region. Given the detection of the \ion{O}{i} $\lambda11287$ line, we would also expect to observe the \ion{O}{i} $\lambda1304$ and \ion{O}{i} $\lambda8446$ transitions, which arise from the same cascade process. The \ion{O}{i} $\lambda1304$ line of the \textit{Forge II} is covered by the DESI spectra, but is not detected ($\leq 2 \sigma$); the \ion{O}{i} $\lambda8446$ is not covered in the current spectra. If resonance fluorescence by Ly$\beta$ is the sole dominant mechanism, the photon flux ratio between \ion{O}{i} $\lambda11287$ and \ion{O}{i} $\lambda1304$ should be unity, corresponding to a flux ratio of 8.7. In this case, the \ion{O}{i} $\lambda1304$ line would have been detected in the DESI spectrum with a S/N of about 44. Several effects can significantly alter this ratio, such as strong UV reddening\cite{Grandi_1983,Rodriguez_2002_OI}. In addition, Ref.\citep{Grandi_1983} describes three processes that can destroy \ion{O}{i} $\lambda1304$ photons: (1) \ion{O}{i} $\lambda1304$ photons can photoionize H~\textsc{i} atoms in the $n=2$ state, (2) collisional de-excitation can destroy \ion{O}{i} $\lambda1304$ photons, and (3) the upper 3s~$^3$S$^0$ level can decay to the metastable terms of the ground configuration via the intercombination lines \ion{O}{i]} $\lambda1641$ and \ion{O}{i]} $\lambda2324$. In fact, their calculations show that up to half of the \ion{O}{i} $\lambda1304$ photons can be converted into \ion{O}{i]} $\lambda1641$\citep{Grandi_1983}. 
The \textit{Forge II} was also observed by the LATIS Survey\cite{2025arXiv251008815N}, in which the $R\sim1000$ IMACS spectra (12-hour integrations) cover 3890--5830~\AA, corresponding to 991--1485~\AA\ in the rest frame of the \textit{Forge II}. The LATIS data independently confirm the absence of any detectable \ion{O}{i} $\lambda1304$ emission, consistent with the DESI result.

\medskip
\noindent {\bf Image decomposition} 

The JWST NIRCam images reveal that the two sources are compact in the long-wavelength channels and extended  in the short-wavelength channels. To quantitatively characterize their morphology, we perform multi-band image decompostion with \texttt{GalfitM} \cite{2013MNRAS.430..330H,2013MNRAS.435..623V} following the standard routine\cite{2023NatAs...7.1376Z, 2025NatAs...}. The decomposition is performed in six NIRCam bands (F115W, F150W, F200W, F277W, F356W, and F444W). Contaminant sources are manually masked to ensure that the fitting process focuses on the central targets. Using \texttt{PSFEx} \cite{2011ASPC..442..435B} and following the strategy in Ref.\cite{2024ApJ...962..139Z}, we construct empirical PSF models by stacking bright, isolated, and unsaturated point sources in the local mosaic region around our targets. Both sources are initially modeled with a PSF component for the AGN and a single Sérsic component for the host galaxy. After the initial fitting process, both sources exhibit substantial off-center residuals in the model-subtracted F200W and F277W images, corresponding to the wavelengths with strong nebular emission lines such as [\ion{O}{iii}]/H$\beta$ and H$\alpha$. Therefore, we decide to include an additional Sérsic component in all bands. 

In the main fitting process, we first use the python codes \texttt{statmorph}\cite{2019MNRAS.483.4140R} and \texttt{photutils}\cite{2024zndo..12585239B} to estimate the position, morphology, and magnitudes for the point source and host galaxy components. The derived values are fed to \texttt{GalfitM} as input parameters. The PSF position of each source is constrained to be within ±1 pixel around its peak pixel (xpeak, ypeak) to ensure that the PSF is fitted to the peak of our targets. For the two Sérsic components, the positions of one Sérsic have initial offsets from the (xpeak, ypeak) to better fit the offset fluxes. To minimize possible degeneracy, we constrain the Sérsic index n to be less than 8 and the half-light radius to be larger than 4.
We perform two iterations of the GalfitM fitting to improve the fitting results. After the first iteration, we use the PSF-subtracted images to refine the morphological model of the host galaxy. The derived parameters serve as inputs for the second iteration.

Extended Data Figures~\ref{fig:galfitm_ID1} and  \ref{fig:galfitm_ID2} show the multi-band image decomposition results. The photometric and morphological parameters are listed in Extended Data Table~\ref{tab:galfitm_res_both}. The morphological parameters of Sérsic components, including the effective radius $R_e$, the Sérsic index $n$, and the positional offset from the PSF center, are the median values measured across all bands. Because a single Sérsic component provides an imperfect description of the complex off-center extended emission, we subsequently perform aperture photometry at the location of the second Sérsic component on the residual image Data$-$PSF$-$Sérsic 1. The nebular flux used in the following SED modeling is taken from these aperture measurements. The fitting results confirm that in the short-wavelength bands ($\lambda_{\rm obs} \leq 2\,\mu \rm m$, corresponding to rest-frame $\lambda_{\rm rest} \lesssim 5000$\,\AA), the emission is dominated by the extended S\'ersic components, indicating that the host galaxy contributes the majority of the flux in the rest-frame optical and UV regime. In contrast, the long-wavelength bands ($\lambda_{\rm obs} > 2\,\mu \rm m$, corresponding to rest-frame $\lambda_{\rm rest} > 5000$\,\AA) exhibit an enhanced point-source contribution from the AGN.

\medskip
\noindent {\bf SED modeling} 

After performing the two-dimensional decomposition of the NIRCam images with \texttt{GalfitM}, we separate the host galaxy, off-center emission, and the central AGN components, and then model them independently. A significant fraction of LRDs exhibit off-center and extended emission. In a systematic study of 99 photometrically selected LRDs at $z \approx 4{-}8$, about $30\%$ were found to exhibit a variety of complex, extended UV morphologies beyond the dominant central point sources\cite{Rinaldi2025}. It is proposed that such off-center emission could arise from nebular emission powered by the central AGN\cite{Chen2025b}, rather than by ionization from young stars. 
For the \textit{Forges}, attributing all the extended emission to the host galaxy would yield an unrealistically high stellar mass, exceeding $10^{12}\,M_\odot$ (see also Z. Li et al., submitted).
This suggests that at least part of the extended light is unlikely to be stellar in origin. Consistent with this, the off-center emission is particularly strong in the F200W and F277W bands, corresponding to the wavelength ranges of [\ion{O}{iii}]/H$\beta$ and H$\alpha$, indicating a potentially non-stellar source of ionization. 
We therefore model the off-center extended emission (Data$-$PSF$-$Sérsic 1) in the NIRCam images (Table~\ref{tab:galfitm_res_both}) as gas clouds photoionized by AGN. We compute a grid of spectral templates using the spectral synthesis code \texttt{CLOUDY}\citep{2023RMxAA..59..327C}.  An open-angle geometry with the inner and stop radii are set to be $R_{\rm d} - R_{\rm e}$ and $R_{\rm d} + R_{\rm e}$, respectively, where $R_{\rm d}$ is the distance from the central point source. For the input ionizing spectrum, we adopt a typical AGN intrinsic UV continuum ($T = 10^6\rm\,K$, $\alpha_{\rm ox}= -1.4$, $\alpha_{\rm UV}=-0.5$, and $\alpha_{\rm X} = -1$, see Section 6.2 of Hazy1\footnote{\url{https://gitlab.nublado.org/cloudy/cloudy/-/tree/master/docs/latex?ref_type=heads}}) with the total AGN luminosity $L_{\rm AGN}$ treated as a free parameter. We assume that a fraction of the UV radiation from the central AGN escapes through a small opening angle of the gas envelope and ionizes the nebula, while the rest is absorbed by the envelope. This fraction is called a covering factor. We derive the covering factor from the solid angle subtended by the off-center \sersic\ component as seen from the PSF component, using its effective radius and projected distance from the PSF component. We then compute spectra with different combinations of total AGN luminosity ($\log [L_{\rm AGN}/{\rm (erg\,s^{-1})}] =  44$--$48$ in steps of 0.5 dex), gas density ($\log[n/{\rm cm^{-3}}] = 1$--$5$ in steps of 1 dex), and metallicity ($Z$ = 0.005, 0.02, 0.2, 0.4, and 1 $Z_\odot$), and apply the intergalactic medium absorption to the output.
We perform Least-squares fitting to identify nebular parameters that best reproduce the observed SEDs of the secondary Sérsic component. The best-fit nebular parameters are $\log [L_{\rm AGN}/{\rm (erg\,s^{-1})}] = 45.5$, $\log[n/{\rm cm^{-3}}] = 4$, and $Z = 1.0\,Z_\odot$ for \textit{Forge I}, and $\log [L_{\rm AGN}/{\rm (erg\,s^{-1})}] = 47.0$, $\log[n/{\rm cm^{-3}}] = 5$, and $Z = 0.4\,Z_\odot$ for \textit{Forge II}.

As shown in Extended Data Figure~\ref{fig:SED}, the decomposition of the NIRCam images reveals a clear decrease in the point-source contribution from the rest-frame near-IR towards the optical–UV bands. We assume that the same trend holds in the UV bands, where the limited image resolution prevents reliable decomposition. These UV bands are dominated by the extended components instead of the central PSF. For the lower-resolution HSC and CFHT images, we derive the galaxy flux by subtracting the modeled nebular emission from the total observed flux, and include these measurements in the SED fitting procedure together with the NIRCam bands. We fit the host galaxies with the SED fitting code \texttt{CIGALE} (v2025; \cite{2019A&A...622A.103B}). We adopt a delayed star formation history with a possible bursty, Bruzual and Charlot (2003) stellar population synthesis models \cite{2003MNRAS.344.1000B}, nebular emission, and a modified Calzetti dust attenuation law \cite{2000ApJ...533..682C}. We include a 2175~\AA\ UV bump to the attenuation curve $A(\lambda)/E(B-V)$, modeled with a Lorentzian-like Drude profile characterized by three parameters: the central wavelength, FWHM, and amplitude. The FWHM and amplitude are allowed to vary during the fitting. The dust emission is modeled with the Draine et al. (2014)\cite{Draine_2014} framework, which treats dust as a mixture of silicate, graphite, and PAHs illuminated by a radiation field ranging from $U_{\min}$ to $U_{\max}$ following a power-law index $\alpha$. The PAH mass fraction ($q_{\mathrm{PAH}}$) and the dust mass fraction associated with star-forming regions ($\gamma$) are also treated as free parameters. To account for uncertainties from the image decomposition, we add a relative error of 5\% in quadrature to all  photometric flux uncertainties. Non-detections in Spitzer MIPS 70~$\mu$m and 160~$\mu$m (and in ALMA \cite{Liu_2019_ALMACS} for \textit{Forge~II}) are included as upper limits. As shown in Extended Data Figure~\ref{fig:SED}, the dust emission from the host galaxies dominates at the rest-frame $\sim 10\ \mu$m, while additional hot dust is required to explain the emission at $\sim 1$–$6\ \mu$m, which cannot be produced by the galaxy alone and is therefore likely contributed by the central AGN.

For the central AGNs, besides the decomposed NIRCam photometry, we also include JWST MIRI F770W, F1280W, F2100W, and Spitzer MIPS 24~$\mu$m photometry \cite{Sanders_2007_SCOSMOS} in the SED model. The AGN flux in the MIRI and Spitzer MIPS bands is calculated by subtracting the modeled galaxy dust emission from the total observed flux. We adopt the geometrically thick, quasi-spherical super-Eddington accretion model proposed by Ref.\cite{LiuH_2025} to interpret the unique SED of the LRDs. To reproduce the hot dust emission at $\sim 1$–$6\ \mu$m, we introduce a blackbody component with a characteristic temperature of $T \sim 500$\,K.

The SED modeling results are shown in Extended Data Figure~\ref{fig:SED}, and the inferred key parameters are summarized in Table~\ref{tab:bh_properties}. The characteristic V-shape observed in the total SEDs is well-modeled by nebular and stellar-dominated emission in the rest-frame UV and LRD gas envelope-dominated emission in the rest-frame optical to near-IR. The bolometric luminosities of the central AGNs are estimated by integrating the reprocessed SEDs of the central point sources, under the assumption that their contribution is negligible in the rest-frame UV due to absorption and reprocessing by the gas envelope. 

The intrinsic UV budget of the AGNs can be estimated from the \texttt{CLOUDY} fitting result of the nebular emission, which is consistent with the observed optical–IR reprocessed emission from the perspective of energy conservation. Previous studies have shown that the dust-to-gas ratio in the nuclear regions of AGNs is highly variable\cite{2021ApJ...906...21J}.
Assuming a Galactic dust-to-gas ratio of $\log [E(B-V)/N_{\rm H}\ (\rm{mag}\ \rm{cm}^2)] = -21.8$, we estimate $E(B-V)$ values from $N_{\rm H}$ derived through X-ray spectral fitting. Applying this $E(B-V)$ and the Calzetti dust extinction law\citep{2000ApJ...533..682C} to the intrinsic UV SED of the AGN with normalization derived from the \texttt{CLOUDY} fitting of the nebular emission, we obtain the attenuated UV SED shown in Extended Data Figure~\ref{fig:Nebular_fit}. As expected, the UV emission from the central AGN is almost completely absorbed, with the absorbed energy consistent with the observed optical–IR reprocessed emission. This implies that, assuming a typical dust-to-gas ratio for the obscuring envelope, X-ray photons from the AGN can still penetrate and be detected, while the UV photons remain trapped and are re-emitted as the observed optical–IR emission. We note that the dense atmospheric gas itself also traps UV emission due to the large optical depth \citep{LiuH_2025}, therefore the real dust-to-gas ratio may be smaller than what we have applied.

\medskip
\noindent {\bf X-ray data reduction}

The \textit{Forges} both have multiple XMM-Newton observations during 2004--2006 from the XMM-Newton Wide-Field Survey in the COSMOS Field \cite{2007ApJS..172...29H}. They also have archival Chandra observations at 2007 from the Chandra Cosmos Legacy Survey \cite{2016ApJ...819...62C}. These data enable study of their X-ray variability. We reduced the raw data of these observations following the method in Ref.\cite{2025ApJ...983...36Z}. 

For the \xmm data reduction, we processed the \hbox{X-ray} data using the \xmm Science Analysis System \citep[SAS v.20.0.0][]{2004ASPC..314..759G} and the latest calibration files. All the EPIC pn, MOS1, and MOS2 data were used in our study when available. We reduced the pn and MOS data following the standard procedure described in the SAS Data Analysis Threads.\footnote{\url{https://www.cosmos.esa.int/web/XMM-Newton/sas-threads}.} Background flares were filtered to generate cleaned event files. For most of the observations, our sources are detected in both the pn and MOS images. In a few observations, the target may be outside the field of view of one or two detectors. For each observation of each source, a source spectrum was extracted using a circular region with a radius of 30$^{\prime\prime}$ centered on the optical source position. A corresponding background spectrum was extracted from a few nearby circular source-free regions on the same CCD chip with a total area of about four times the area of the source region. Spectral response files were generated using the {\sc rmfgen} and {\sc arfgen} tasks. We grouped the source spectra with at least one count per bin, since the total spectral counts numbers in the 0.3--10 keV band combining the pn and MOS spectra for all observations are smaller than 300\cite{2025ApJ...983...36Z}. This results in 6 epochs and 5 epochs of \xmm spectra for \textit{Forge~I} and \textit{Forge~II}, respectively. We also stacked the XMM-Newton spectra of each source to achieve higher S/N.

To reduce the Chandra data, we used the Chandra Interactive Analysis of Observations (CIAO; v4.15)\footnote{\url{http://cxc.harvard.edu/ciao/}.} tool. We first used the CIAO {\sc  chandra\_repro} script to create new bad-pixel files and new level 2 event files. Background flares were removed using the {\sc deflare} script with an iterative 3$\sigma$ clipping algorithm. Then a source spectrum was extracted using the {\sc specextract} tool from a circular region centered on the optical position. The radius of the circle was chosen to enclose $90\%$ of the PSF at 1 keV using the {\sc psf} command. A background spectrum was extracted from an annular region centered on the source position, with the inner and outer radii chosen to be the source radius plus 4$^{\prime\prime}$ and 8$^{\prime\prime}$, respectively. We have visually inspected the background-extraction regions and veriﬁed that they did not contain any \hbox{X-ray} sources. We combined the extracted Chandra spectra of each source using {\sc combine\_spectra} since they were all observed within several days. Chandra spectra were also grouped with at least one count per bin.

The \hbox{X-ray} spectra of were fitted using XSPEC\cite{arnaud1996astronomical} (v12.12.1) with the $W$ statistic (\hbox{$W$-stat}).\footnote{\url{https://heasarc.gsfc.nasa.gov/xanadu/xspec/manual/XSappendixStatistics.html} for details.} In order to check the variability of these two sources, we first adopted a simple \hbox{power-law} model modified by Galactic absorption ({\sc zpowerlw*phabs}) to describe the multi-epoch 0.3--10 keV \xmm spectra or \hbox{0.5--7 keV} Chandra spectra. For the XMM-Newton observation, we jointly fitted the available EPIC pn, MOS1, and MOS2 spectra. We also added normalization constants (\hbox{best-fit} values between 0.7 and 1.3) to the MOS spectra to account for small cross-calibration uncertainties. The simple \hbox{power-law} model can describe all the spectra well. The combined XMM-Newton spectra are shown in Figure~\ref{fig:Xray_spec}. For \textit{Forge~II}, a Fe K$\alpha$ line at a rest-frame energy of 6.4~keV is detected with a confidence level of $98.4\%$ ($p = 0.016$; estimated using the {{\sc lrt} command of \texttt{XSPEC}) and a moderately large rest-frame EW of 0.71~keV, indicating the presence of dense, cold material that reflects hard X-rays in the vicinity of the nucleus. For \textit{Forge~I}, we did not detect significant Fe K$\alpha$ line (confidence level of $62.7 \%$, $p = 0.373$) due to its lower luminosity. From the best-fit power-law models, we computed the rest-frame 2 keV flux densities ($f_{\rm 2keV}$), 10 keV flux densities ($f_{\rm 10keV}$), and \hbox{2--10~keV} luminosities ($L_{\rm X}$) of each source in each epoch. The derived X-ray light curves of the two sources are shown in Figure~\ref{fig:Xray_lc}, with the black dashed lines and gray shaded regions showing the fitted values and errors of the stacked XMM-Newton spectra using the same model. Both \textit{Forge~I} and \textit{Forge~II} display some levels of X-ray variability. \textit{Forge~I} exhibits significant variability with a maximum amplitude of $5.4^{+37.8}_{-2.7}$ in $L_{2-10 \rm keV}$ over a timescale of $\sim 2$ years in the observed frame, whereas \textit{ForgeII} shows a modest variability amplitude of $1.7^{+0.8}_{-1.0}$ over $\sim 0.5$ years. To characterize their absorption level, we fitted the stacked XMM-Newton spectra with an intrinsic absorption component ({\sc zphabs}), and the photon index of the intrinsic power-law was a free parameter. The best-fit $N_{\rm H}$ are $2.1^{+1.8}_{-1.3} \times 10^{22} \,\rm{cm}^{-2}$ and $1.4^{+1.8}_{-1.3} \times 10^{22}\, \rm{cm}^{-2}$, with intrinsic photon indices of $2.4^{+0.4}_{-0.5}$ and $1.5^{+0.3}_{-0.2}$. These column densities indicate a modest level of obscuration.

\medskip

\noindent {\bf The $\alpha_{\rm ox}-L_{2500}$ relation and Fundamental plane} \\
The UV–to–X-ray power-law spectral slope, $\alpha_{\rm ox}$, is a widely used indicator of the strength of ionizing emission in quasars. Large quasar samples have revealed a tight correlation between $\alpha_{\rm ox}$ and the rest-frame 2500~\AA\ luminosity, $L_{2500}$. We estimate the intrinsic $L_{2500}$ of the central AGNs by modeling their ionized nebular emission (see SED modeling). As shown in the left panel of Fig.~\ref{fig:aox_FP}, the \textit{Forges} lie on the $\alpha_{\rm ox}$–$L_{2500}$ relation defined by the general quasar population, indicating that their central AGNs are approaching the typical quasar regime. In contrast, the observed total $L_{2500}$ values fall beyond the $3\sigma$ scatter of the relation, suggesting that the observed UV emission is dominated by the host galaxy and nebular components rather than the AGN itself.

Another well-established relation is the fundamental plane (FP), which links the radio luminosity, X-ray luminosity, and black hole mass of accreting black holes. It encapsulates the coupling between the accretion disc and the jet, unifying X-ray binaries and active galactic nuclei under a common disc–jet framework across mass scales. Radio-quiet and radio-loud AGNs occupy distinct FPs, with the former exhibiting systematically higher Eddington ratios than the latter. As shown in the middle and right panels of Figure~\ref{fig:aox_FP}, the \textit{Forges} lie on the FP of radio-quiet AGNs, consistent with their radio loudness values of $R<10$. This further supports the interpretation that their central AGNs are approaching the typical quasar regime.

\medskip
\noindent {\bf UV variability} \\
JWST Cycle~1 COSMOS-Web program\cite{Cweb_survey_ref} and the COSMOS-3D program obtained NIRCam images in F115W and enable us to probe the rest-frame 3000~\AA\ variability on a rest-frame timescale of 200 days. We performed F115W-band photometry at the two epochs using a small aperture radius of 0\farcs15 to isolate the variability of the central region. For \textit{Forge~I}, we obtain $26.15 \pm 0.37$ and $26.54 \pm 0.43$ mag in the two epochs, respectively. For  \textit{Forge~II}, the corresponding values are $24.61 \pm 0.20$ and $24.68 \pm 0.21$ mag. The total photometric uncertainty includes contributions from the source Poisson noise, the background noise within the aperture, and the uncertainty in the background estimation from the sky annulus, and is expressed as
\[\sigma_\mathrm{flux}=\sqrt{\mathrm{flux}+N_{\mathrm{aperture}} \sigma_B^2+\frac{N_{\mathrm{aperture}}^2 \sigma_B^2}{N_{\mathrm{annual}}}},
\]
where $N_{\mathrm{aperture}}$ and $N_{\mathrm{annual}}$ are the numbers of pixels within the source aperture and sky annulus, respectively, and $\sigma_B$ is the standard deviation of the background within the annulus. 
The variability significance between the two epochs is quantified as
\[
\mathrm{SNR}_{\mathrm{var}}=\frac{|\Delta \mathrm{flux}|}{\sqrt{\sigma_{\mathrm{flux,Epoch1}}^{2}+\sigma_{\mathrm{flux,Epoch2}}^{2}}}.
\]
This estimation yields low $\mathrm{SNR}_{\mathrm{var}}$ values of 0.68 and 0.26 for the two \textit{Forges}, indicating no apparent UV variability.   

\medskip
\noindent {\bf Number density} \\
The wavelength coverage of the F444W slitless spectra from COSMOS-3D enables detection of \ion{He}{i} at redshifts $z=2.5$–$3.7$. We identify all \ion{He}{i}-emitting LRDs within the available COSMOS-3D area (9/10 of the whole survey observed as of November 2025; effective coverage of $\sim$0.28 deg$^{2}$) to estimate the fraction of LRDs caught in the transition phase at this epoch. 
Our parent sample originates from the COSMOS-Web photometric catalog\citep{2025arXiv250603243S} and includes only those objects with detected emission lines of S/N$>3$ in the COSMOS-3D coverage.
We select compact object with $f_{0\farcs3}/f_{0\farcs15} < 1.8 $.  
For each object, we fit the \ion{He}{i} emission line with one narrow Gaussian component (FWHM $<$ 500 $\rm km~s^{-1}$) or one broad Gaussian component (FWHM $>$ 1000 $\rm km~s^{-1}$) plus one narrow Gaussian component (FWHM $<$ 500 $\rm km~s^{-1}$). 
To compare the quality of these two types of fitting, we calculate the Bayesian Information Criterion (BIC) parameter, defined as BIC=$\chi^2+k\ln n$, where $k$ is the number of free parameters and $n$ is the number of data points. 
We select broad-line objects with $\rm{BIC}_{2Gaussian} - \rm{BIC}_{1Gaussian} <5$ and a broad-component S/N larger than 5.
We visually inspect all the broad-line objects to remove the contamination from nearby objects and identify the \ion{He}{i} lines based on the photometric redshift and multiple emission lines around \ion{He}{i}. 
We then measure their UV/optical continuum slopes and selected V-shape objects following the criteria of Ref.~\citep{Kocevski2025}.
This procedure yields seven LRDs (including \textit{Forges}) with $M_{5100} = -$25.2 to $-$20.1, implying that $\sim$30\% of \ion{He}{i}-emitting LRDs at $z\sim2.5$–3.7 appear to be in the transition stage. The evolutionary state of the remaining population is still uncertain. Interestingly, the other five LRDs show no X-ray or radio detections and lack elevated MIR emission, consistent with the properties of typical LRDs. Only one of them exhibits a blueshifted \ion{He}{I} absorption line, but that feature is much weaker than in \textit{Forges}'. These contrast between \textit{Forges} and the other five \ion{He}{I} LRDs further supporting the uniqueness of \textit{Forges}.

We put the inferred number density of \ion{He}{i} LRDs and transitioning LRDs in the context of number density evolution of LRDs and quasars in Extended Data Figure~\ref{fig:density_evolution}. The error bars of our estimated densities represent Poisson uncertainties. For comparison, we adopt the quasar number density from Ref.\citep{2020MNRAS.495.3252S} with $M_{\rm UV}< -21$, which corresponds to $M_{5100} < -20.5$ for a typical quasar SED.  
Our estimated LRD number densities at $z\sim2.5$–3.7 are smaller than Ref.\citep{Ma2025} and broadly consistent with the prediction of the log-normal model\citep{Inayoshi2025_firstactivity}. The measurements from Ref.\citep{Ma2025} are based solely on photometry and do not account for the presence of broad spectral lines, and may therefore include some contamination. At this redshift range, the LRD number density is comparable to that of quasars, while transitioning LRDs constitute a small fraction of the population, in good agreement with the transition scenario we proposed.


\clearpage


\renewcommand{\figurename}{Extended Data Figure}
\renewcommand{\tablename}{Extended Data Table}

\setcounter{figure}{0}
\setcounter{table}{0}

\begin{table}[H] 
        \centering
        \caption{Emission and Absorption Line Properties. All EWs are in the rest frame. Velocities are relative to the systemic redshift, with positive values indicating redshifted and negative values indicating blueshifted.}
        \label{tab:linefits}
        \begin{tabular}{lllll}
        \hline
        Feature & Component & Parameter & \textit{Forge~I} & \textit{Forge~II} \\
        \hline
        &  & Redshift  & 2.869 & 2.925 \\
        \hline
        \multicolumn{5}{c}{\textbf{Emission Lines}} \\
        \hline
        \ion{He}{i} & Component 1    & FWHM (km s$^{-1}$) & $8352.8 \pm 12.9$ & $6771.9 \pm 65.4$ \\
                                                    &               & EW (\AA)          & $134.4 \pm 7.1$  & $97.0 \pm 5.9$  \\
                                                    & Component 2  & FWHM (km s$^{-1}$) & $1764.4 \pm 20.1$  & $2069.9 \pm 23.3$  \\
                                                    &              & EW (\AA)           & $44.8 \pm 4.3$  & $59.9 \pm 5.1$  \\
        Pa$\gamma$                       &               & FWHM (km s$^{-1}$) & $3254.4 \pm 16.6$  & $4726.4 \pm 24.8$ \\
                                                    &               & EW (\AA)           & $37.3 \pm 3.2$  & $70.2 \pm 3.3$  \\
        Pa$\delta$                           &                & FWHM (km s$^{-1}$) & \textemdash  & $4726.2 \pm 22.8$ \\
                                                    &                & EW (\AA)           &  \textemdash  & $46.9 \pm 2.0$  \\
        \ion{O}{i} $\lambda11287$  &                & FWHM (km s$^{-1}$) & $3437.3 \pm 59.6$   & $3701.9 \pm 58.3$  \\
                                                    &                & EW (\AA)           & $31.8 \pm 3.2$   & $20.5 \pm 2.5$   \\
        \hline
        \multicolumn{5}{c}{\textbf{Absorption Features}} \\
        \hline
        \ion{He}{i}             & Component 1 & $v_{\rm shift}$ (km s$^{-1}$) & $      35.7 \pm 45.0$   & $-18.9 \pm 34.1$ \\
                                                    &            & $\tau_0$                      & $0.38 \pm 0.06$    & $0.85 \pm 0.06$  \\

                                                    &            & $\sigma$ (km s$^{-1}$)         & $279.1 \pm 44.2$    & $214.4 \pm 16.9$  \\
                                                    &            & $\log{N}$ (cm$^{-2}$)         & $13.24 \pm 0.13$    & $13.47 \pm 0.06$  \\

                                                    & Component 2 & $v_{\rm shift}$ (km s$^{-1}$) & $300.4 \pm 36.9$  & $472.2 \pm 26.8$ \\
                                                    &            & $\tau_0$                      & $0.66 \pm 0.08$    & $1.59 \pm 0.07$  \\
                                                    &            & $\sigma$ (km s$^{-1}$)         & $96.4 \pm 14.5$    & $203.1 \pm 8.7$  \\
                                                    &            & $\log{N}$ (cm$^{-2}$)         & $13.01 \pm 0.09$    & $13.72 \pm 0.04$  \\

        Pa$\delta$                          &             & $v_{\rm shift}$ (km s$^{-1}$) & \textemdash  & $-317.5 \pm 107.0$ \\
                                                    &            & $\tau_0$                      & \textemdash    & $0.07 \pm 0.03$  \\
                                                    &            & $\sigma$ (km s$^{-1}$)         & \textemdash    & $125.0 \pm 56.0$  \\
                                                    &            & $\log{N}$ (cm$^{-2}$)         & \textemdash    & $13.40 \pm 2.53$  \\

        \hline
        \multicolumn{5}{c}{\textbf{UV Spectrum}} \\
        \hline
        Ly$\alpha$                             &            & FWHM (km s$^{-1}$) & \textemdash & $556.4 \pm 32.7$ \\
                                                &         & EW$_{\rm obs}$ (\AA) &  \textemdash   & $102.7 \pm 5.2$ \\

                                                &            &$v_{\rm shift}$ (km s$^{-1}$)         & \textemdash  & $ 443.2 \pm  13.9$ \\
        \hline
        \end{tabular}
        \end{table}    

                           
\begin{table}[H]
\centering
\caption{Photometry and Morphological Parameters of the \textit{Forge~I} and \textit{Forge~II}.}
\label{tab:galfitm_res_both}
\begin{tabular}{lcccc}
\hline
Parameter & PSF & Sérsic 1 & Sérsic 2 & Data$-$PSF$-$Sérsic 1 \\
\hline
\multicolumn{5}{c}{\textit{Forge~I}} \\
\hline
$R_e$ (kpc)        & -- & $0.98 \pm 0.01$ & $2.88 \pm 0.10$ &  -- \\
Sérsic $n$         & -- & $0.66 \pm 0.03$ & $0.80 \pm 0.05$ &  -- \\
offset (kpc)       & -- & $0.46 \pm 0.01$ & $3.73 \pm 0.10$ &  -- \\
F115W       & $26.46 \pm 0.05$ & $25.98 \pm 0.05$ & $25.95 \pm 0.10$ & $25.82 \pm 0.33$ \\
F150W       & $25.40 \pm 0.05$ & $26.16 \pm 0.08$ & $26.78 \pm 0.32$ & $24.82 \pm 0.19$ \\
F200W       & $23.44 \pm 0.05$ & $24.41 \pm 0.05$ & $23.53 \pm 0.05$ & $23.44 \pm 0.05$ \\
F277W       & $21.63 \pm 0.05$ & $23.27 \pm 0.05$ & $23.87 \pm 0.05$ & $23.55 \pm 0.05$ \\
F356W       & $21.16 \pm 0.05$ & $22.68 \pm 0.05$ & $24.72 \pm 0.05$ & $24.02 \pm 0.05$ \\
F444W       & $20.36 \pm 0.05$ & $22.18 \pm 0.05$ & $25.43 \pm 0.09$ & $23.67 \pm 0.05$ \\
\hline
\multicolumn{5}{c}{\textit{Forge~II}} \\
\hline
$R_e$ (kpc)        & -- & $0.95 \pm 0.03$ & $2.81 \pm 0.16$ &  -- \\
Sérsic $n$         & -- & $4.19 \pm 0.08$ & $1.18 \pm 0.08$ &  -- \\
offset (kpc)       & -- & $0.30 \pm 0.03$ & $1.95 \pm 0.16$ &  -- \\
F115W       & $25.68 \pm 0.05$ & $24.49 \pm 0.10$ & $24.90 \pm 0.13$ & $24.59 \pm 0.11$ \\
F150W       & $24.57 \pm 0.05$ & $23.96 \pm 0.06$ & $24.80 \pm 0.15$ & $23.79 \pm 0.13$ \\
F200W       & $23.14 \pm 0.05$ & $22.70 \pm 0.05$ & $23.08 \pm 0.04$ & $22.94 \pm 0.05$ \\
F277W       & $21.34 \pm 0.05$ & $21.46 \pm 0.05$ & $23.85 \pm 0.07$ & $22.88 \pm 0.05$ \\
F356W       & $20.70 \pm 0.05$ & $20.93 \pm 0.05$ & $24.48 \pm 0.15$ & $23.89 \pm 0.05$ \\
F444W       & $19.90 \pm 0.05$ & $20.58 \pm 0.05$ & $25.49 \pm 0.23$ & $22.49 \pm 0.05$ \\
\hline
\end{tabular}
\end{table}

\begin{figure}[H]
    \centering
    \includegraphics[width=\linewidth]{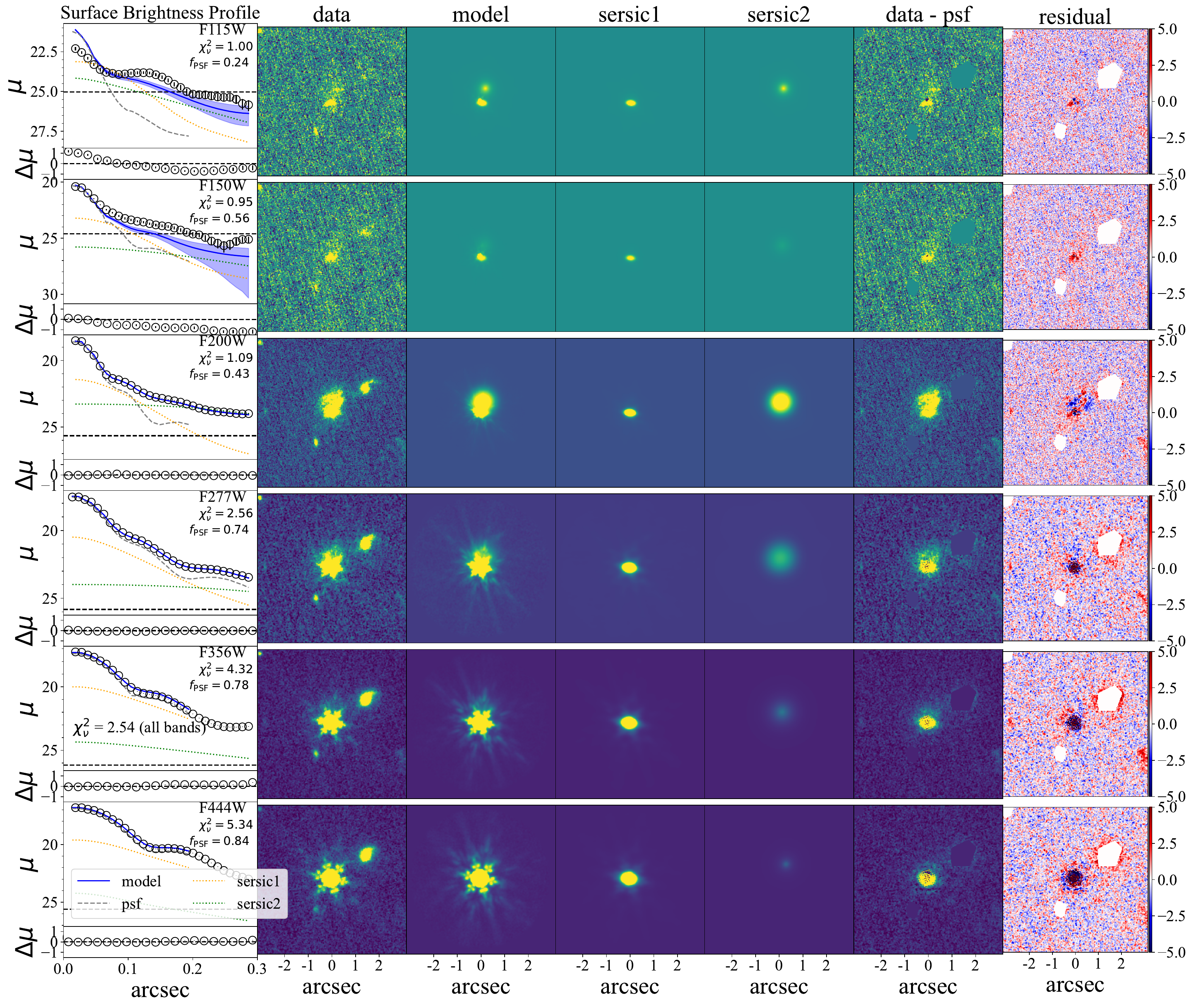}
    \caption{Two-dimensional image decomposition of the \textit{Forge~I} with \texttt{galfitM}. 
    For each JWST/NIRCam band (F115W–F444W), panels from left to right show the surface-brightness ($\mu$, in the units of $\rm mag\,arcsec^{-2}$) profiles, data, best-fit model, individual Sérsic components, data - PSF, and residuals/error that are stretched linearly from $-$5 to 5. The reduced $\chi^2$ values and the PSF flux fractions ($f_{\rm PSF}$) are indicated for each band. The $\chi^2$ value for all eight bands is given in the lower-left corner of the panel for F356M.}
    \label{fig:galfitm_ID1}
\end{figure}  

\begin{figure}[H]
    \centering
    \includegraphics[width=\linewidth]{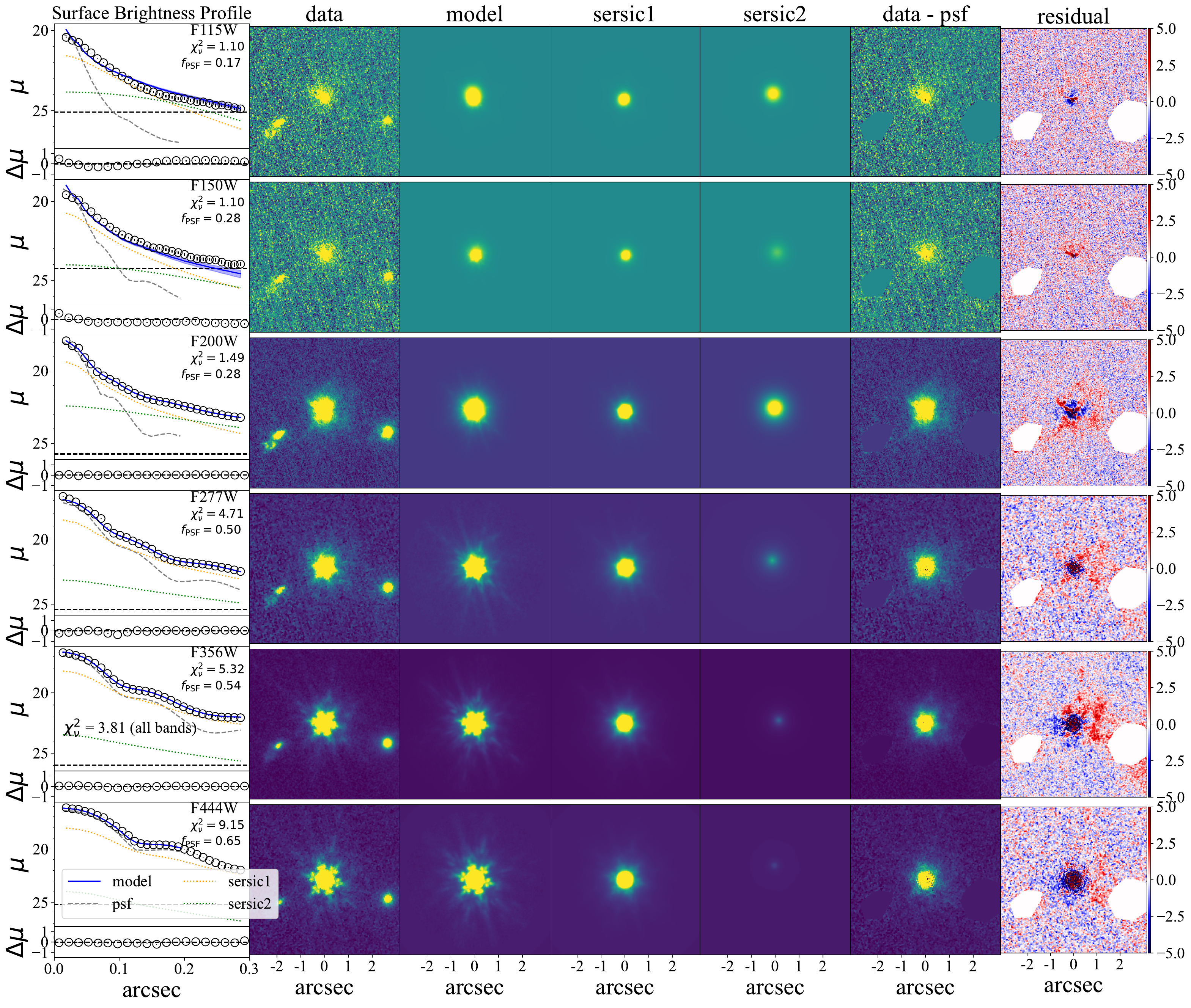}
    \caption{Same as Figure above, but for the \textit{Forge~II}.}
    \label{fig:galfitm_ID2}
\end{figure} 

\begin{figure}[H]  
    \centering
    \begin{subfigure}{0.49\linewidth}
        \centering
        \includegraphics[width=\linewidth]{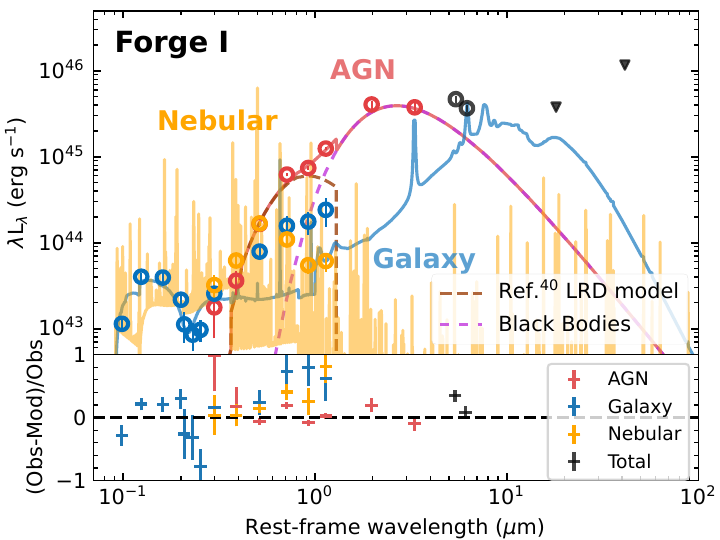}
        \label{fig:SED_a}
    \end{subfigure}
    \begin{subfigure}{0.49\linewidth}
        \centering
        \includegraphics[width=\linewidth]{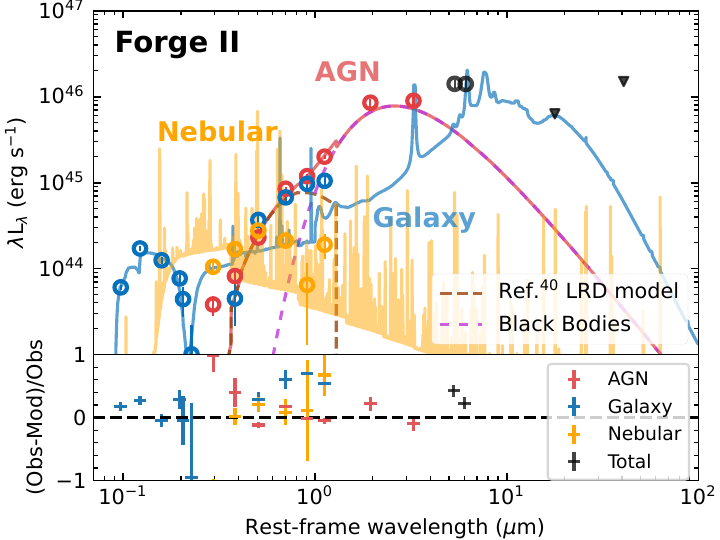}
        \label{fig:SED_b}
    \end{subfigure}
    \caption{Best-fit SED modeling. In each figure, the top panels show the observed photometry (open circles with error bars; upper limits are marked by downward triangles) together with the best-fit models (solid curves). The photometry is decomposed into AGN, host-galaxy, and nebular components in the rest-frame UV–near-IR, whereas only the total fluxes or upper limits are presented in the mid- to far-IR because of the limited spatial resolution. The AGN model consists of an LRD model with optically thick photospheres around a super-Eddington accreting black hole\cite{LiuH_2025}, combined with one blackbody at characteristic temperatures of $T \sim 500$\,K. The bottom panel shows the residuals, defined as (Observed $-$ Model)/Observed, for the AGN, galaxy and nebular components separately.}
    \label{fig:SED}
\end{figure}    

\begin{figure}[H]
    \centering
    \includegraphics[width=\linewidth]{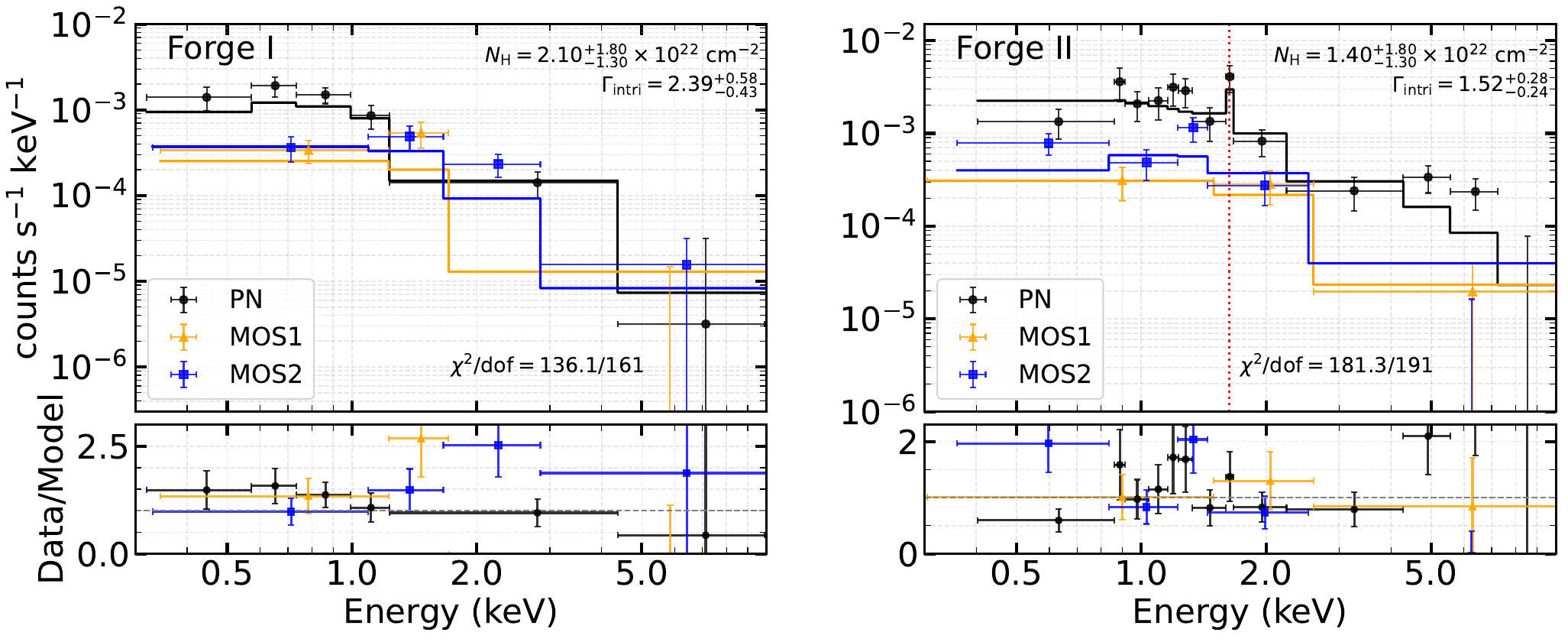}
    \caption{The combined \hbox{XMM-Newton} spectra of the \textit{Forges} overlaid with the best-fit \hbox{power-law+zphabs} models. The bottom panels show the ratios of the spectral data to the best-fit models. The EPIC pn (black), MOS1 (orange), and MOS2 (blue) spectra were jointly fitted. For \textit{Forge~II}, a Fe K$\alpha$ line at a rest-frame energy of 6.4~keV (observed-frame 1.63~keV) is detected, with its position indicated by the red dashed line.}
    \label{fig:Xray_spec}
\end{figure} 

\begin{figure}[H]
    \centering
    \includegraphics[width=\linewidth]{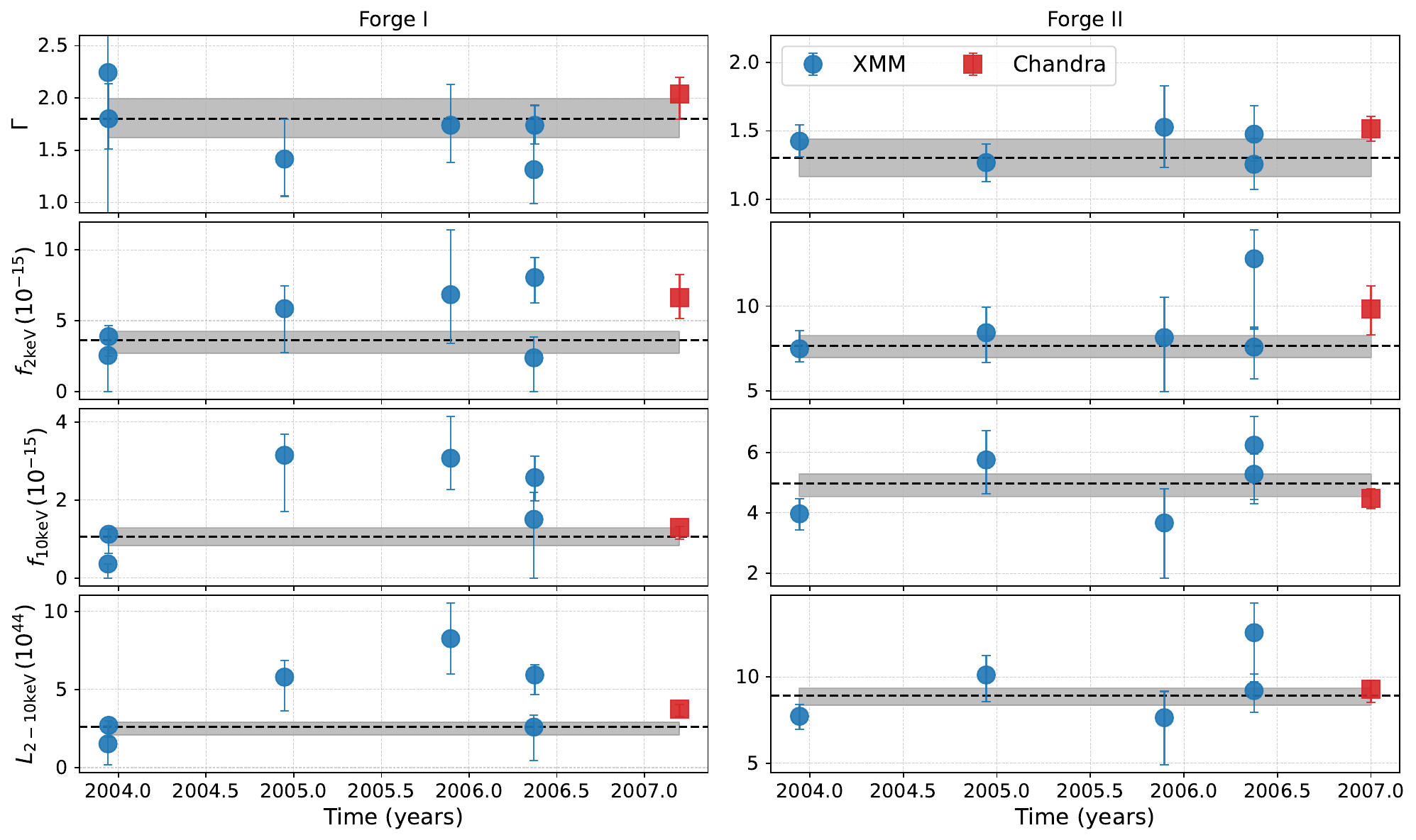}
    \caption{Variability of the photon index ($\Gamma$), X-ray flux densities at rest-frame 2~keV and 10~keV, and the 2–10~keV X-ray luminosity. The black dashed lines and gray shaded regions represent the measured values and errors of the stacked XMM-Newton spectra.}
    \label{fig:Xray_lc}
\end{figure} 

\begin{figure}[H]
    \centering
    \includegraphics[width=0.49\linewidth]{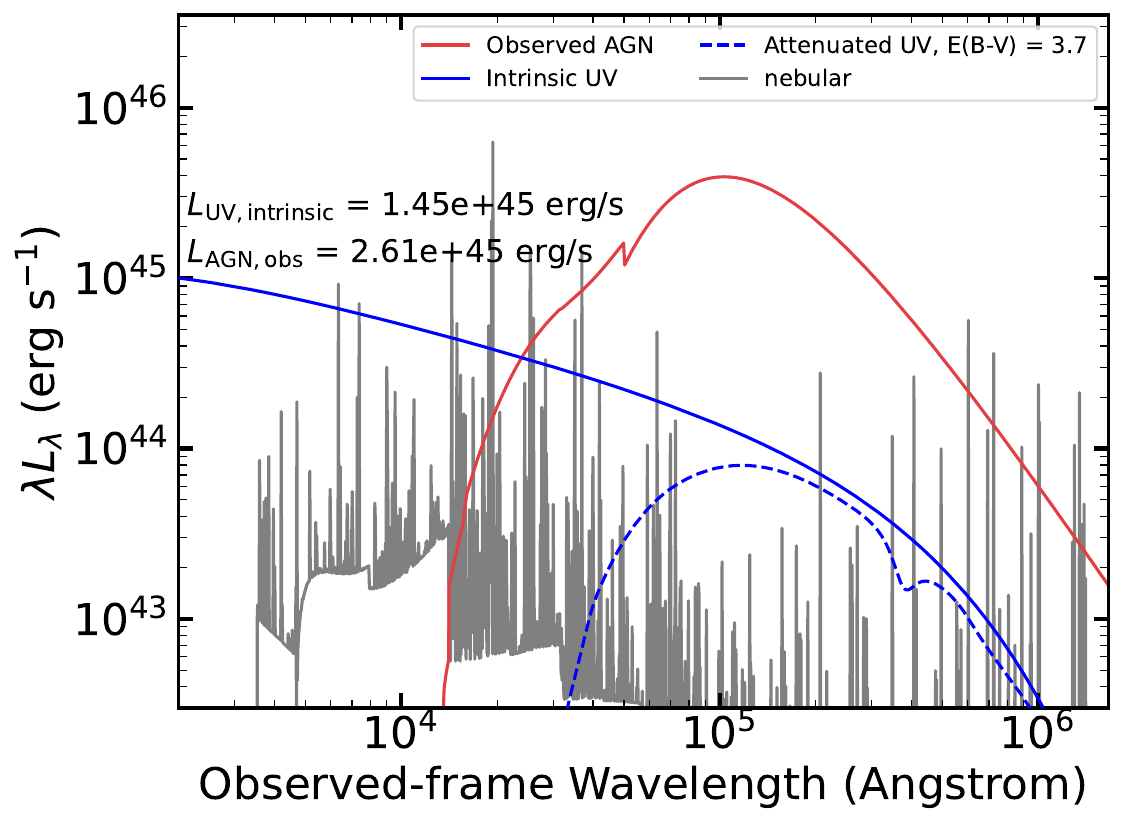}
    \includegraphics[width=0.49\linewidth]{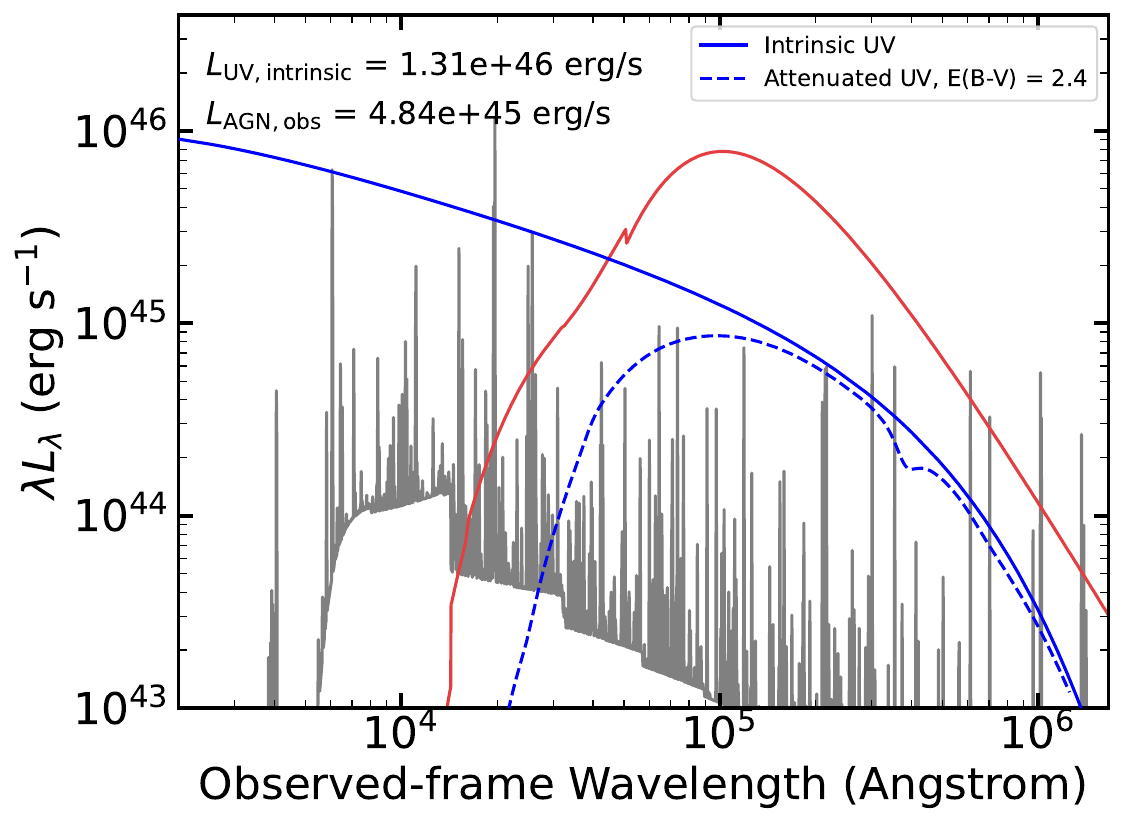}
    \caption{The intrinsic UV SED of the central AGN inferred from \texttt{CLOUDY} fitting of the ionized nebular emission of \textit{Forge I} (left) and \textit{Forge II} (right), compared with the attenuated UV SED assuming the Galactic dust-to-gas ratio and Calzetti dust extinction law\citep{2000ApJ...533..682C}. The gas column density $N_{\rm H}$ is derived from X-ray spectral fitting. The absorbed UV energy and the observed reprocessed optical–IR energy are indicated, showing comparable magnitudes.}
    \label{fig:Nebular_fit}
\end{figure} 

\begin{figure}[H]
    \centering
    \includegraphics[width=\linewidth]{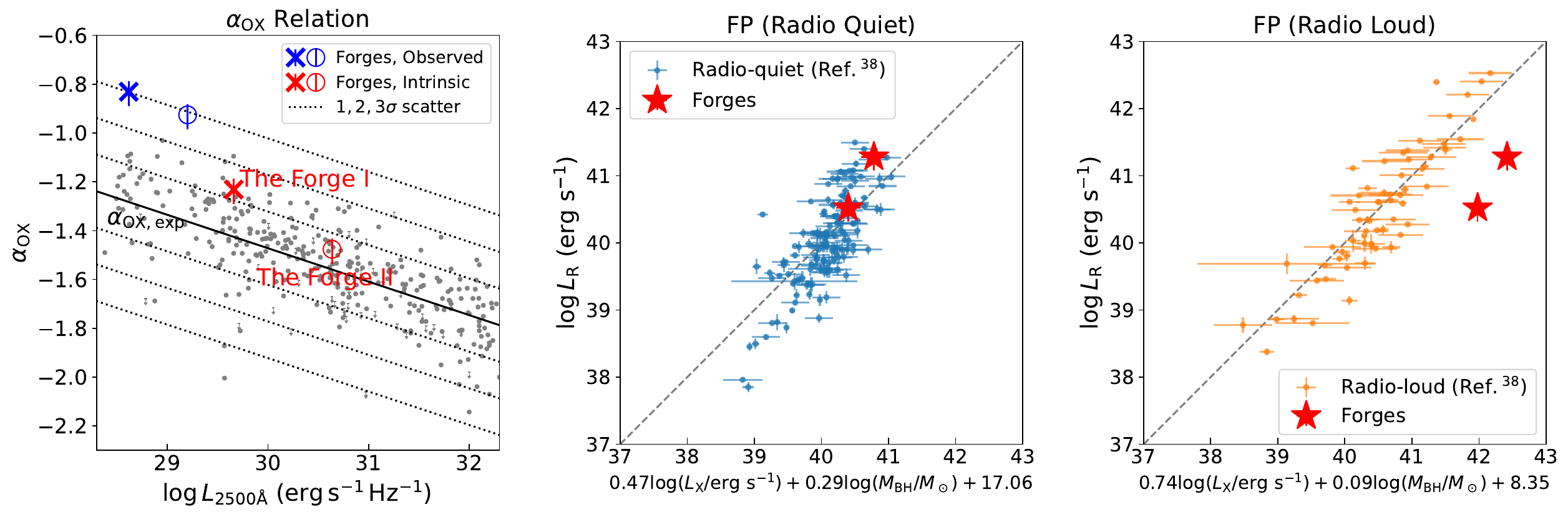}
    \caption{\textbf{Left:} Locations of the \textit{Forges} on the $\alpha_{\rm ox}$–$L_{2500}$ plane. Blue symbols show the observed total $L_{2500}$, which is dominated by host-galaxy and off-center nebular emission, while red symbols indicate the intrinsic $L_{2500}$ inferred from the ionized nebular emission. The $\alpha_{\rm OX} \textrm{--} L_{\rm 2500 \AA}$ relation from Ref.~\cite{2006AJ....131.2826S} is shown as the solid line, with the dotted lines representing the $1,\,2,\,3 \sigma$ scatter of the relation. The small grey dots and downward arrows represent the $\alpha_{\rm OX}$ values and upper limits of the quasar sample of Ref.~\cite{2006AJ....131.2826S}, respectively.
    \textbf{Middle and Right:} Locations of the \textit{Forges} on the fundamental plane\cite{2024A&A...689A.327W}, where $L_{\rm R}$ is the rest-frame 5~GHz radio luminosity (in units of $\rm{erg\, s^{-1}}$), $L_{\rm X}$ is the rest-frame 2–10~keV X-ray luminosity (in units of $\rm{erg\ s^{-1}}$), and $M_{\rm BH}$ is the black hole mass (in units of $M_\odot$), shown for radio-quiet AGNs (middle panel) and radio-loud AGNs (right panel). The dashed black line in each panel represents the 1:1 relation. }
    \label{fig:aox_FP}
\end{figure}

\begin{figure}[H]
    \centering
    \includegraphics[width=0.55\linewidth]{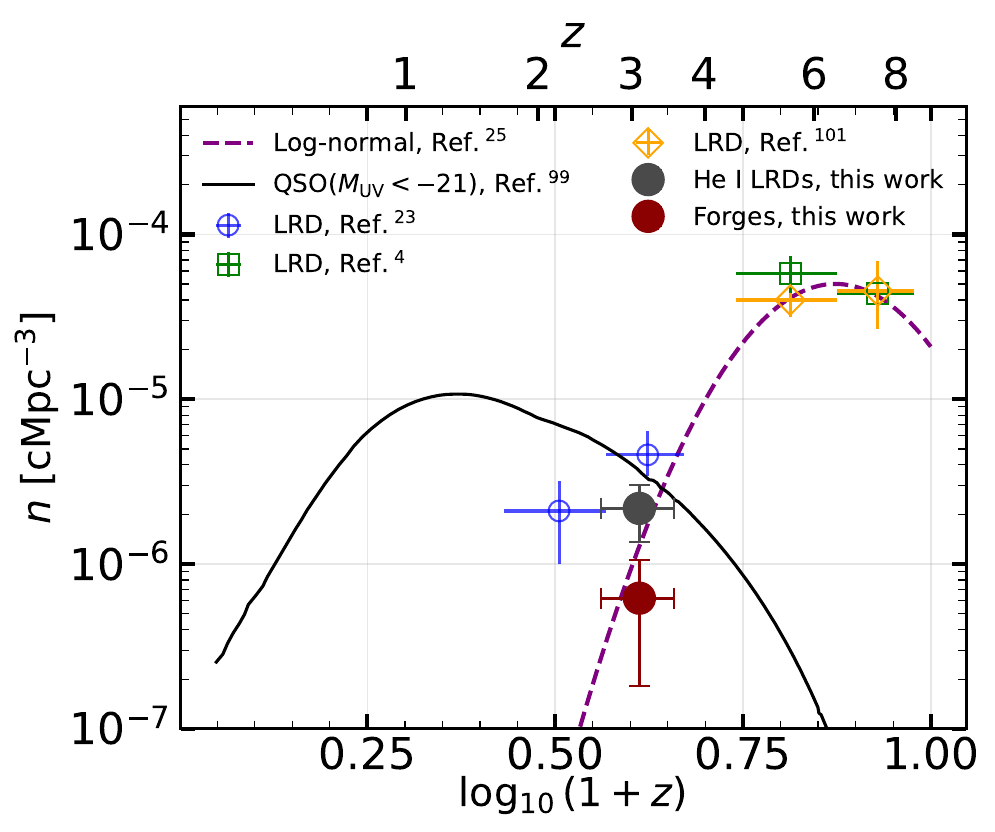}
    \caption{Number density evolution for LRDs with $M_{5500} < -20.5$. The solid gray and red dots show the \ion{He}{I} LRDs and transitioning LRDs at $z = 2.5$–$3.7$ estimated in the COSMOS-3D footprint, whose error bars represent Poisson uncertainties. These are compared with literature measurements\citep{Kokorev2024,Kocevski2025,Ma2025}, a log-normal LRD evolution model\citep{Inayoshi2025_firstactivity} (dashed purple line), and the quasar number density evolution with $M_{\rm UV}<-21$ from Ref.\citep{2020MNRAS.495.3252S} (solid black curve).}
    \label{fig:density_evolution}
\end{figure} 


\newpage
\section*{Data Availability}
All JWST data and HST slitless spectrum data are publicly available at MAST: \url{https://archive.stsci.edu/}. Reduced HST F814W image are downloaded from \url{https://irsa.ipac.caltech.edu/data/COSMOS/images/acs_mosaic_2.0/}. HSC images can be downloaded from \url{https://hsc-release.mtk.nao.ac.jp/doc/index.php/data-access__pdr3/}. CFHT image can be downloaded from \url{http://www.asiaa.sinica.edu.tw/~whwang/musubi}. XMM-Newton data are publicly available at \xmm Science Archive: \url{https://www.cosmos.esa.int/web/xmm-newton/xsa}. Chandra data are publicly available at: \url{https://cxc.cfa.harvard.edu/cda/}.

\section*{Code Availability}
Image reduction and analysis used publicly available pipelines: CEERS NIRCam \cite{2023ApJ...946L..12B}, 
CIAO\footnote{\url{https://cxc.cfa.harvard.edu/ciao/}.}, 
CIGALE\cite{2019A&A...622A.103B}, 
CLOUDY\citep{2023RMxAA..59..327C},
GalfitM \cite{2013MNRAS.430..330H,2013MNRAS.435..623V}, 
GRIZLI\footnote{\url{https://grizli.readthedocs.io/en/latest/}.},
PSFEx\cite{2011ASPC..442..435B}, PyPHER\cite{2016ascl.soft09022B}, SAS\cite{2004ASPC..314..759G}, 
Source Extractor \cite{1996A&AS..117..393B}, XSPEC\cite{arnaud1996astronomical}.  

\section*{Acknowledgements}
We acknowledge support from the National Science Foundation of China (12225301). L.C.H. was supported by the China Manned Space Program (CMS-CSST-2025-A09) and the National Science Foundation of China (12233001). K.I. acknowledges support from the National Natural Science Foundation of China (12573015, 1251101148) and the Beijing Natural Science Foundation (IS25003). We thank Changhao Chen, Ruancun Li, Junfeng Wang, Zijian Li, and Siwei Zou for helpful discussions.

\section*{Author Contributions}
S.F. and Z.Z. analyzed the data, performed the calculations and wrote the manuscript. D.J. and J.C. contributed to the data analysis and manuscript preparation. L.J. helped design the project and write the manuscript. K.C. contributed to the reduction of the spectral data. L.C.H., K.I., J.L., F.S., F.W. and J.Y. contributed to the interpretation of the data and to the final manuscript.

\section*{Competing Interests Statement} 
The authors declare that they have no competing interests.

\bibliography{ms} 

\end{document}